%% file: gmos-kinematics.tex
\documentclass[twocolumn,times]{aastex62}

\newcommand{\kms}{km s$^{-1}$} 

\graphicspath{{./}{plots/}{scratch/}}

\received{XXXX}
\revised{YYYY}
\accepted{ZZZZ}

\shorttitle{MASSIVE XIII -- GMOS Kinematics}
\shortauthors{Ene et al.}

\usepackage[utf8]{inputenc}
\usepackage{graphicx}
\usepackage{amssymb}
\usepackage{amsmath}
\usepackage{natbib}
\begin{document}

\title{The MASSIVE Survey XIII -- Spatially Resolved Stellar Kinematics in the Central 1 kpc of 20 Massive Elliptical Galaxies with the GMOS-North Integral-Field Spectrograph}

\correspondingauthor{Irina Ene}
\email{irina.ene@berkeley.edu}

\author{Irina Ene}
\affiliation{Department of Astronomy, University of California, Berkeley, CA 94720, USA}
\affiliation{Department of Physics, University of California, Berkeley, CA 94720, USA}

\author{Chung-Pei Ma}
\affiliation{Department of Astronomy, University of California, Berkeley, CA 94720, USA}
\affiliation{Department of Physics, University of California, Berkeley, CA 94720, USA}

\author{Nicholas J. McConnell}
\affiliation{University of California, 1156 High Street, Santa Cruz, CA 95064, USA}

\author{Jonelle L. Walsh}
\affiliation{George P. and Cynthia Woods Mitchell Institute for Fundamental Physics and Astronomy, and Department of Physics and Astronomy, \\
Texas A\&M University, College Station, TX 77843, USA}

\author{Philipp Kempski}
\affiliation{Department of Astronomy, University of California, Berkeley, CA 94720, USA}

\author{Jenny E. Greene}
\affiliation{Department of Astrophysical Sciences, Princeton University, Princeton, NJ 08544, USA}

\author{Jens Thomas}
\affiliation{Max Plank-Institute for Extraterrestrial Physics, Giessenbachstr. 1, D-85741 Garching, Germany}

\author{John P. Blakeslee}
\affiliation{Gemini Observatory, Casilla 603, La Serena, Chile}



\begin{abstract}

We use observations from the GEMINI-N/GMOS integral-field spectrograph (IFS) to obtain spatially resolved stellar kinematics of the central $\sim 1$ kpc of 20 early-type galaxies (ETGs) with stellar masses greater than $10^{11.7} M_\odot$ in the MASSIVE survey.
Together with observations from the wide-field Mitchell IFS at McDonald Observatory in our earlier work, we obtain unprecedentedly detailed kinematic maps of local massive ETGs, covering a scale of $\sim 0.1-30$ kpc.
The high ($\sim 120$) signal-to-noise of the GMOS spectra enable us to obtain two-dimensional maps of the line-of-sight velocity, velocity dispersion $\sigma$, as well as the skewness $h_3$ and kurtosis $h_4$ of the stellar velocity distributions.
All but one galaxy in the sample have $\sigma(R)$ profiles that increase towards the center, whereas the slope of $\sigma(R)$ at one effective radius ($R_e$) can be of either sign.  
The $h_4$ is generally positive, with 14 of the 20 galaxies having positive $h_4$ within the GMOS aperture and 18 having positive $h_4$ within $1 R_e$.  
The positive $h_4$ and rising $\sigma(R)$ towards small radii are indicative of a central black hole and velocity anisotropy.
We demonstrate the constraining power of the data on the mass distributions in ETGs by applying Jeans anisotropic modeling (JAM) to NGC~1453, the most regular fast rotator in the sample.  
Despite the limitations of JAM, we obtain a clear $\chi^2$ minimum in black hole mass, stellar mass-to-light ratio, velocity anisotropy parameters, and the circular velocity of the dark matter halo.

\end{abstract}

\keywords{galaxies: elliptical and lenticular, cD --- galaxies: evolution --- galaxies: formation --- galaxies: kinematics and dynamics --- galaxies: structure}



\section{Introduction}

As the final product of multiple merger events, massive early-type galaxies (ETGs) in the local universe provide excellent insight into how galaxies evolve.
ETGs are complex, multi-component systems, and a full description of their evolutionary processes must take into account the stars, dark matter, supermassive black holes (SMBHs), and any gas present in the galaxies.

Significant recent progress on understanding local ETGs has been made by surveys using integral field spectrographs (IFS), e.g., SAURON \citep{Emsellem2004}, ATLAS$^{\rm 3D}$ \citep{cappellarietal2011}, SAMI \citep{croometal2012}, CALIFA \citep{sanchezetal2012}, and MaNGA \citep{Bundyetal2015}.
These surveys use wide-field IFS to produce two-dimensional maps of the stellar and gas kinematics
and investigate fundamental galaxy properties such as the dichotomy between fast and slow rotators, early-type galaxy morphologies, and molecular and ionized gas content.
The ETGs targeted by these surveys are predominantly fast-rotating S0 or elliptical galaxies with $M_* < 10^{11.5} M_\odot$.  The spatial sampling of these survey is limited by the IFS fiber diameter ($1.6''$, $2''$, $2.7''$ for SAMI, MaNGA, CALIFA, respectively) or lenslet size ($0.94''$ for SAURON/ATLAS$^{\rm 3D}$).  
A few other recent IFS or long-slit studies of a smaller number of ETGs specifically targeted brightest cluster galaxies
(BCGs; e.g. \citealt{Broughetal2011, jimmy13, Loubseretal2018}),
and the SLUGGS survey used multi-slits to reach a sky coverage of $\sim 2 - 4$ effective radii for a subset of ATLAS$^{\rm 3D}$ galaxies \citep{brodieetal2014}.

We initiated the volume-limited MASSIVE survey (\citealt{Maetal2014}) to investigate
the $\sim 100$ most massive galaxies located up to a distance of 108 Mpc in the northern sky. 
The survey targets a complete sample of ETGs with an absolute $K$-band magnitude brighter than $M_K = -25.3$ mag, or a stellar mass greater than $M_* \sim 10^{11.5} M_\odot$, a parameter space little explored previously. 
Wide-field kinematics and stellar population studies from this survey have been published in \cite{Greeneetal2015}, \cite{Pandyaetal2017}, \citet{Vealeetal2017a, Vealeetal2017b, Vealeetal2018}, \cite{Eneetal2018}, and \cite{Greeneetal2019}
based on IFS data from the Mitchell Spectrograph on the 2.7-m Harlan J. Smith Telescope at McDonald Observatory.
These two-dimensional kinematics have a spatial resolution of $\sim 4''$ (fiber diameter) and cover a field of view (FOV) of $107'' \times 107''$.
Paper V (\citealt{Vealeetal2017a})
found a dramatic increase in the fraction of slow-rotating ETGs with increasing $M_*$, reaching $\sim 90$\% at $M_* \sim 10^{12} M_\odot$.
Paper VII \citep{Vealeetal2017b} examined the relationship of galaxy spin, $M_*$ and environment and
found that galaxy mass, rather than environment, is the primary driver of the apparent kinematic morphology-density relation for local ETGs.  The physical processes responsible for building up the present-day stellar masses of massive ETGs must be very efficient at reducing their spin, in any environment.
Paper VIII \citep{Vealeetal2018} investigated the environmental dependence of the stellar velocity dispersion profiles and found the fraction of galaxies with rising outer profiles to increase with $M_*$ and environmental density, a trend likely due to total mass variations rather than velocity anisotropies.
Paper X (\citealt{Eneetal2018}) analyzed substructures in the stellar velocity maps and found kinematic twists and large scale ($R \gtrsim 10''$) kinematically distinct components.

In this paper (Part XIII), we present the first results from the high angular-resolution spectroscopic portion of the MASSIVE survey.  While the earlier wide-field IFS studies offered insight to galaxies' global dynamics and assembly histories, kinematics of the innermost regions of galaxies are critical for measuring the masses of the SMBHs and for elucidating the symbiotic relations among black holes, baryons and dark matter near galactic centers.
To achieve these goals, 
we observed the central $5'' \times 7''$ of 20 MASSIVE galaxies using Gemini Multi Object Spectrograph (GMOS, \citealt{Hooketal2004}) with $0.2''$ lenslets on the 8.1 m Gemini North Telescope.  The exposure times were chosen to yield stellar spectra with signal-to-noise ratio (S/N) of $\sim 120$ for each spatial bin.
These high-quality spectra enable us to
obtain detailed two-dimensional maps of the velocity, velocity dispersion, as well as the skewness and kurtosis of the stellar line-of-sight velocity distribution (LOSVD).  Depending on the galaxy, the maps contains from 50 to 300 spatial bins (with an average of 130 bins) and cover a scale from $\sim 0.1$ kpc to a few kpc.

The size of the sample and the IFS spatial coverage in this paper is similar to that of \cite{McDermidetal2006}, who studied the central $8'' \times 10''$ region of 28 ETGs in the SAURON survey using the OASIS spectrograph with a spatial sampling of $0.27''$.
The fine spatial sampling enabled them to identify two types of kinematically distinct components (KDCs) in lower-mass ETGs: old, kpc-scale KDCs that exist in slow rotating ETGs, and young, small ($\sim 100$ pc scale), almost counter-rotating KDCs in fast rotators.
However, their kinematics were obtained from spectra with a lower S/N of 60.
Their sample is in the $M_*$ range of $10^{10} - 10^{11.6} M_\odot$, where 68\% are fast rotators and many show emission lines. By contrast, the galaxies studied here are mostly slow rotators and few have emission lines. There is no overlapping galaxy with the two samples

The remaining sections of the paper are organized as follows.
In Section~\ref{sec:galaxy_sample_and_data} we present the sample of 20 MASSIVE galaxies with high resolution IFS observations.
In Section~\ref{sec:gmos_n_observations} we describe the observations and the data reduction pipeline.
Section~\ref{sec:measuring_stellar_kinematics} provides an overview of how we use the IFS data sets to measure stellar kinematics.
Sections~\ref{sec:velocity_features},~\ref{sec:velocity_dispersion}, and~\ref{sec:higher_order_moments} examine the behavior of the stellar kinematics in the galaxies' nuclei:
Section~\ref{sec:velocity_features} explores the velocity profiles, 
Section~\ref{sec:velocity_dispersion} looks at the radial behavior of velocity dispersion, 
and Section~\ref{sec:higher_order_moments} studies the higher order moments $h_3$ and $h_4$.
In Section~\ref{sec:jam_model} we showcase how the combined set of small and large scale kinematics can be use to constrain dynamical models of the fast rotator NGC~1453.
Section~\ref{sec:conclusions} summarizes our main conclusions.


\section{Galaxy sample}
\label{sec:galaxy_sample_and_data}

An in-depth description of the selection of the parent sample of 116 galaxies is given in \citet{Maetal2014}.
In summary, MASSIVE is a volume-limited survey of the most luminous early-type galaxies with $M_K \lesssim -25.3$ mag (from 2MASS; \citealt{skrutskieetal2006}) corresponding to $M_* \gtrsim 10^{11.5} M_\odot$, that are within a distance of 108 Mpc in the northern sky (declination $\delta > -6\degr$).

\begin{deluxetable*}{lrcrccccccrcr}
\tablecaption{Galaxy properties and GMOS observational details for the 20 MASSIVE galaxies. \label{tab:galaxy_props}}
\tablehead{
\colhead{Galaxy} & \colhead{$D$} & \colhead{$M_K$} & \colhead{PA$_{\rm phot}$} & \colhead{$R_e$} & \colhead{$\lambda_e$} &
\colhead{$\sigma_e$} & \colhead{$\lambda_{\rm 1 kpc}$} & \colhead{$\sigma_{\rm 1 kpc}$} & \colhead{$\gamma_{\rm 1 kpc}$} & \colhead{IFU PA} & \colhead{Semester} & \colhead{Exposure} \\
\colhead{} & \colhead{[Mpc]} & \colhead{} & \colhead{[deg]} & \colhead{[$''$ (kpc)]} & \colhead{} & \colhead{[km s$^{-1}$]} & \colhead{} & \colhead{[km s$^{-1}$]} & \colhead{} & \colhead{[deg]} & \colhead{} & \colhead{}
}
\colnumbers
\input{galaxy_properties}
\tablecomments{
    (1) Galaxy name.
    (2) Distance from Paper I \citep{Maetal2014}.
    (3) Absolute $K$-band magnitude from Paper I \citep{Maetal2014}.
    (4) Photometric position angle, taken from Paper IX \citep{Goullaudetal2018}. *Values for NGC~2340 and NGC~4874 are from 2MASS and NSA, respectively.
    (5) Effective radius from CFHT deep $K$-band photometry (Quenneville et al, in preparation).
    (6) Spin parameter within a circular aperture of radius $1 R_e$ from Paper X \citep{Eneetal2018}. 
    (7) Luminosity-weighted average velocity dispersion within a radius of $1 R_e$ (from column 5), measured from the Mitchell IFS.
    (8) Spin parameter within a radius of 1 kpc measured from the GMOS IFS.
    (9) Luminosity-weighted average velocity dispersion within a radius of 1 kpc measured from GMOS data.
    (10) Power law slope of $\sigma(R)$ measured from GMOS data. See Sec.~\ref{sub:velocity_dispersion_radial_profiles} for definition.
    (11) Position angle of the long axis of the GMOS IFU.
    (12) Semester when the GMOS data were taken.
    (13) Science exposure times.}
\end{deluxetable*}

In this paper we present results for 20 galaxies that were chosen for follow-up observations with high spatial resolution spectroscopy with GMOS. 
The key properties of these galaxies are summarized in Table~\ref{tab:galaxy_props}.
The 20 galaxies are located at a distance in the range of 54.4 Mpc to 102.0 Mpc (with a median distance of $\sim 70$ Mpc) and have $-25.50~{\rm mag} \geq M_K \geq -26.33~{\rm mag}$, which corresponds to stellar mass $10^{11.7} M_\odot \lesssim M_* \lesssim 10^{12} M_\odot$.
Since they were selected based on the ability to obtain high-$S/N$ GMOS data for dynamical mass modeling,
this sample of 20 galaxies tends to be the closer and more massive part of the general MASSIVE sample: 
they represent $\sim 50\%$ of galaxies within 80 Mpc and $\sim 60\%$ of galaxies more massive than $M_K < -25.8~{\rm mag}$.


\section{GMOS-N observations and data reduction}
\label{sec:gmos_n_observations}

Our galaxies were observed using the GMOS integral field unit (IFU, \citealt{Allington-Smithetal2002}) on the 8.1 m Gemini North telescope.
The observations were taken in queue mode over six semesters between 2012 -- 2016 under the programs GN-2012B-Q-31, GN-2013B-Q-29, GN-2013B-Q-68, GN-2014B-DD-6, GN-2015A-Q-19, GN-2015B-Q-59, and GN-2016B-Q-18.
Total exposure times were chosen to ensure a S/N of $\sim 120$ (after spatial binning) and vary between 1 and 6 hours (see Table~\ref{tab:galaxy_props}).
All galaxies presented here were observed after the GMOS-N upgrade to e2v deep depletion detectors in 2011 \citep{Rothetal2012}, but before the upgrade to Hamamatsu fully-depleted detectors in 2017 \citep{Scharwachteretal2018}.

All observations were taken using the two-slit mode of the GMOS-N IFU. This provides a FOV of $5'' \times 7''$ consisting of 1000 hexagonal lenslets with a projected diameter of $0.2''$.
An additional 500 lenslets observe a $5'' \times 3.5''$ region of the sky displaced by $\sim 1$ arcmin from the science field.
The lenslets are coupled to fibers that map the focal plane into two pseudo-slits (each covering half of the FOV) through which light is passed to the rest of the spectrograph.
Each pseudo-slit covers the same spectral range and is projected across the full spatial dimension of the detector array (perpendicular to the dispersion direction).  
In the spectral dimension of the array, the two pseudo-slits are projected with an offset in the central wavelength and thereby avoid overlap if the spectral range is sufficiently narrow.
We use the R400-G5305 grating + CaT filter combination to avoid spectral overlap on the detector.
This results in a clean wavelength range of 7800 \AA\ -- 9330 \AA\ that has good coverage of the CaII triplet and NaI absorption features used for measuring stellar kinematics and stellar populations, respectively.

The detector array consists of three 2k $\times$ 4k e2v deep depletion CCDs placed in a row with $\sim 37$ unbinned pixels gaps in between.
To mitigate the loss of spectral information to the chip gaps, we use spectral dithering with a grating central wavelength $\lambda_c$ for half of the exposures and $\lambda_c + 50$ \AA~for the other half.
For most galaxies, a typical value of $\lambda_c$ is between 8600 \AA~and 8700 \AA.
We carefully choose this value to ensure that the CaII triplet lines do not fall on either of the two chip gaps.

The basic reduction of the raw data frames is performed using the Gemini package within the image reduction and analysis facility (IRAF) software.
For an in-depth example of how to reduce GMOS IFU data using IRAF (and potential pitfalls) see \citet{Lena2014}.
We use the standard GMOS data reduction procedure.
The science, flat lamp, and twilight raw frames are bias subtracted.
The arc frames are taken in fast-read mode so they are only overscan subtracted.
The Gemini calibration unit (GCAL) flat lamp frames (taken before and after the science exposures) are used to identify and extract the trace of each fiber on the detector array and to determine the flat-field response map.
The twilight exposure is used to correct for illumination.
The GCAL arc lamp frames are used to determine the wavelength solution by fitting a fourth-order Chebyshev polynomial to known CuAr arc lamp lines spanning the wavelength range of the observations.
The science spectra are extracted using the fiber traces identified in the flat lamp exposures. The spectra are then corrected for flat-fielding and illumination.
Cosmic ray artifacts are removed from the spectra and spectral processing is performed using the arc lamp wavelength solution.
We use a custom routine to perform sky subtraction. For each pseudo-slit, we subtract an average spectrum computed from the dedicated sky fibers corresponding to that particular pseudo-slit.
The end result of this step is a reduced science frame that contains the one-dimensional spectrum corresponding to each GMOS lenslet.

For all galaxies observed in semester 2013B or later, arc lamp exposures were recorded immediately adjacent to science exposures of each galaxy, with the telescope at the same pointing.  For three galaxies observed in semester 2012B (NGC 315, NGC 741, and NGC 2340), arc lamp exposures were recorded in daytime with the telescope parked.  The arc-calibrated science frames from 2012B exhibit residual wavelength errors,
most readily apparent as a sharp wavelength offset between bright sky lines in the two GMOS pseudo-slits.  For each arc-calibrated science frame (prior to sky subtraction), we parameterized the residual wavelength offset $\epsilon_\lambda$ in each GMOS lenslet by fitting $\sim 10$ bright sky lines and fitting a polynomial function $\epsilon_\lambda(\lambda)$ across the observed wavelength range.  To mitigate the low signal-to-noise of sky lines in individual lenslets, we fit $\epsilon_\lambda(\lambda)$ as a first-order polynomial in each lenslet independently.  The resulting two-dimensional map of $\epsilon_\lambda$ in each science frame is interpolated to a wavelength of interest (i.e. centered on the CaII triplet), smoothed using a 20-lenslet boxcar, and applied to a convolution kernel for the stellar template spectrum during kinematic fitting (Section~\ref{sec:measuring_stellar_kinematics}).
These calibration steps simultaneously measure the wavelength- and lenslet-dependent instrumental resolution $\Delta\lambda$.
Since we account for this instrumental term during kinematic fitting and we ultimately measure kinematics from co-added galaxy spectra, the corresponding instrumental kernels for ($\epsilon_\lambda$, $\Delta\lambda$) in individual lenslets are co-added as well.

The individual reduced science exposures are not stacked or mosaicked.
Instead, we use a suite of custom routines to extract and co-add one-dimensional spectra from multiple exposures.
We construct collapsed (along the wavelength direction) images of the galaxy and determine the location of the galactic center by fitting a Moffat profile to the light profiles of the collapsed images.
Then, we extract the one-dimensional spectra and tag each of them by the exposure number and that lenslet's spatial coordinates relative to the galaxy center.
We interpolate the extracted spectra to a common wavelength basis and perform a heliocentric velocity correction if the exposures were recorded over multiple nights.

\begin{figure}
  \includegraphics[width=\columnwidth]{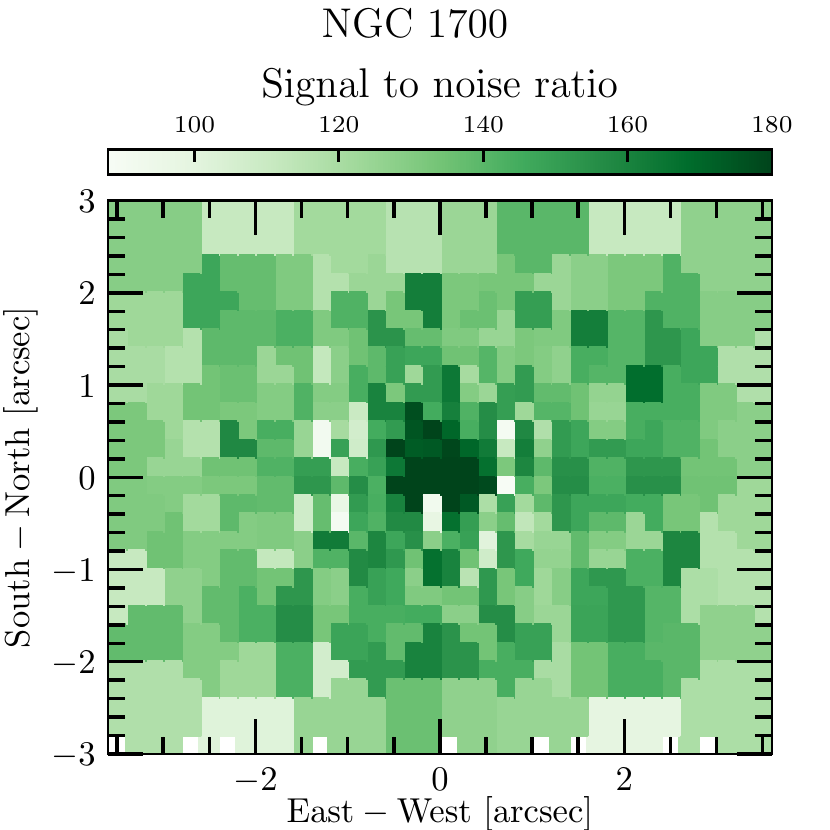}
    \caption{
    Example signal-to-noise ratio map for NGC~1700.
    The S/N value for each Voronoi bin scatters around the target of 125 with a typical RMS scatter of $\sim 10\%$.
    Individual $0.2'' \times 0.2''$ lenslets (each square grid) in the innermost region have S/N much greater than the target value. Stellar kinematics from high-quality spectra in this region are critical for measuring the gravitational influence of any supermassive black hole present at the galaxy's center.
    }
    \label{fig:example-binmap}
\end{figure}

To increase the S/N of the data, we use the Voronoi binning procedure of \citet{CappellariCopin2003}.
The one-dimensional spectra are irregularly spaced in galactocentric coordinates, due to dithers and pointing offsets between individual frames, and the hexagonal shape of the IFU lenslets.
Since Voronoi binning requires regularly spaced coordinates, we construct a square grid in (x,y) with a grid spacing of $0.2''$, equal to the width of a hexagonal lenslet.
We then tag each spectrum to the nearest grid point.
We do not consider overlap with multiple grid points -- each spectrum is given 100\% weight at a single grid position.

The Voronoi binning procedure does not co-add the spectra, but merely defines the bins based on the estimated signal and noise of each point on the regular grid.
For each grid point,  we estimate the fluxes and the pixel-to-pixel variance of all contributing spectra by using the residuals between the observed spectra and boxcar-smoothed spectra over a 10 pixel window.
The signal assigned to each grid point is then the sum of the fluxes of all contributing spectra, while the noise is the quadrature sum of the contributing spectra noise.
We use a custom implementation of the binning step that imposes spatial symmetry over four galaxy quadrants.
The bin boundaries are then used to create symmetric bins in the remaining three quadrants.
The data (i.e. the 1-D spectra) is never folded during this step.
We use a Voronoi binning target S/N of 125, which results in high quality spectra that do not sacrifice the spatial resolution of the innermost bins.
The resulting S/N per bin values generally scatter around the target, with a typical RMS scatter of $\sim 10\%$, as can be seen in Fig.~\ref{fig:example-binmap}.

In most cases, a Voronoi bin is composed of multiple $0.2'' \times 0.2''$ grid segments, each affiliated with multiple one-dimensional spectra, usually from different exposures.
The final step is to co-add all of the one-dimensional spectra in each Voronoi bin and to create a corresponding Gaussian kernel for the instrumental resolution of the co-added spectrum.
This is done with a clipped 3-$\sigma$ mean and rescaling for regions overlapping the chip gaps at one of the two wavelength settings.


\section{Measuring stellar kinematics}
\label{sec:measuring_stellar_kinematics}

The stellar LOSVD is extracted using the penalized pixel-fitting (pPXF) method of \citet{CappellariEmsellem2004}, which convolves a set of template stellar spectra with the LOSVD function $f(v)$.
The latter is modeled as a Gauss-Hermite series of order up to $n=6$ (\citealt{Gerhard1993, vanderMarelFranx1993}):

\begin{equation}
\label{eq:GH}
f(v) = \frac{e^{-\frac{y^2}{2}}}{\sqrt{2 \pi \sigma^2}} \bigg[1 + \sum_{m=3}^n h_m H_m(y) \bigg],
\end{equation}

\noindent where $y=(v-V)/\sigma$, $V$ is the mean velocity, $\sigma$ is the velocity dispersion, and $H_m$ is the $m^{\text{th}}$ Hermite polynomial as defined in Appendix A of \cite{vanderMarelFranx1993}.

For each binned spectrum we fit all six Gauss-Hermite moments, although for brevity we only show the first four moments in our kinematic plots.
We run pPXF with an initial guess of 0 for $V$ and $h_3$ through $h_6$, and 300 km s$^{-1}$ for $\sigma$.
As our model continuum fit, we use an additive polynomial of degree zero (i.e. an additive constant) and a multiplicative polynomial of degree three.
In cases where the LOSVD is undersampled or the S/N is low, it is important to use the pPXF penalty term to suppress the large uncertainties of the higher order moments \citep{CappellariEmsellem2004}.
Since the galaxies in our sample have large velocity dispersions and the data have high S/N, it is not critical that we penalize deviations from a Gaussian solution -- hence we set the pPXF keyword BIAS to zero.

\begin{figure}
    \includegraphics[width=\columnwidth]{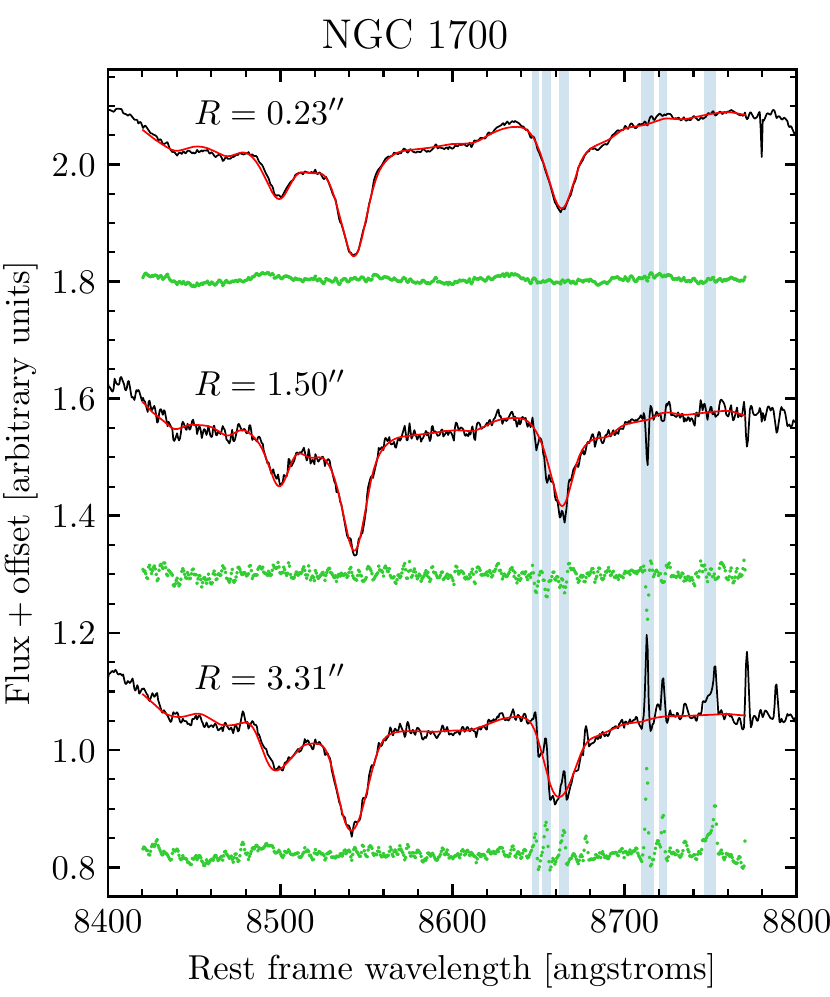}
    \caption{Example fits to the GMOS stellar spectra of NGC~1700 at the galaxy center (top), at an intermediate distance from the center (middle), and in one of the outermost bins (bottom).  The observed spectrum is shown in black and the stellar template broadened by the best-fit LOSVD is overplotted in red. The green dots represent the residuals between data and model and are shifted by an arbitrary amount for clarity. The fit region is centered around the triplet of CaII absorption lines and spans the wavelength range 8420-8770 \AA.
     The light blue shaded regions indicate improperly subtracted sky lines and are excluded from the fit.
    }
    \label{fig:example-spectra}
\end{figure}

Prior to fitting, we center the spectra on the triplet of calcium absorption features by cropping close to the wavelength range 8420 -- 8770 \AA.
We also mask any prominent residual sky lines.
That is, we mask small wavelength regions centered on the locations where the sky lines occur.
Overall, this corresponds to excluding $\sim10\%$ or less of the fit region.
We present example pPXF fits at the nucleus, in an intermediate and in an outer spatial bin for NGC~1700 in  Fig.~\ref{fig:example-spectra}.

\citet{Barthetal2002} tested the robustness of using the Ca triplet spectral region for velocity dispersion measurements. Similar to \cite{Dressler1984}, they found little sensitivity in the measurements to the choice of template stars.  
We performed our own template mismatch tests using 
two different sets of template stars chosen from
the CaT Library of 706 stars in \citet{Cenarroetal2001}.
The first template set contains the identical
15 K and G stars as in Table 2 of \citet{Barthetal2002}.  The second template set contains all $\sim 360$ G and K stars in the CaT Library.  For the latter, we ran pPXF 
for all the bins in one of our 20 galaxies and examined the 40 stars that were assigned the highest weights.
Only one of the 40 stars is in common with the 15 stars in \cite{Barthetal2002}.  Despite the difference in the two template choices,
we find that the rms difference in the kinematic moments measured from the two sets of templates is $\sim 5$ \kms~for $V$ and $\sigma$, and $\sim 0.01$ for $h_3$ and $h_4$, well within the measurement errors of $\sim 10$ \kms~and $\sim 0.02$, respectively. We therefore confirm that our kinematic measurements are robust to template choices.
The results reported in this paper use the 15 template stars in \citet{Barthetal2002}.

We note that we performed similar tests in \cite{Vealeetal2017a} 
for the Mitchell IFU data in the wavelength range $3650 - 5850$ \AA.  The results there also 
indicated that our kinematic measurements are relatively insensitive to the choice of input template library.

The stellar spectra in the CaT Library cover the wavelength range 8348 -- 9020 \AA~with a spectral resolution of 1.5 \AA~FWHM.
To match the spectral resolution of our binned spectra, we convolve the templates with a Gaussian distribution with appropriate dispersion.
We determine the instrumental resolution of our data by using the known sky lines prior to sky subtraction, as described in Section~\ref{sec:gmos_n_observations}, or arc lines from the CuAr calibration lamp.
We first fit a Gaussian to each individual sky or arc line.
Then we fit a low-order polynomial to the full width at half-maximum (FWHM) of the lines versus wavelength and determine the corresponding FWHM for our fit region centered at $\sim 8600$ \AA.
We determine this best-fitting FWHM for each individual lenslet of each exposure.
Typical values for the FWHM are $\sim 2.5$ \AA, with variations of $\sim 0.3$ \AA~with lenslet position.
This corresponds to an instrumental resolution of $\sim 37$ \kms~at 8600 \AA~with a sampling of 0.67 \AA~per pixel.
Finally, we generate a Gaussian kernel for each binned spectrum by averaging the kernels of each individual lenslet assigned to that bin.
While it is possible for the line spread function to deviate mildly from a Gaussian shape (e.g., slightly flat-topped for the KCWI spectrograph; \citealt{Morrisseyetal2018, vanDokkumetal2019}), in practice 
the uncertainty in the LSF introduces non-negligible bias in the recovery of the kinematic moments only when the measured velocity velocity dispersion is smaller than the instrumental dispersion \citep{Cappellari2017}.  Typical velocity dispersions for MASSIVE galaxies are $\ga 250$ \kms, much higher than the spectral resolution of GMOS ($\sim 40$ \kms). Uncertainties in the LSF are therefore subdominant to other sources of systematic error.

The error bars on the kinematic moments are obtained through a bootstrap approach.
This choice is motivated by the fact that each Voronoi bin contains tens (inner bins) to hundreds (outer bins) of individual lenslet spectra from different science exposures that have different noise properties.
A bootstrap trial consists of taking the individual lenslet spectra that belong to a Voronoi bin and drawing a sample with replacement, then using this sample to generate a new co-added spectrum, which we run through pPXF to determine the best-fitting Gauss-Hermite moments.
We repeat this process 100 times.
Finally, the error for each kinematic moment is computed as the standard deviation of the pPXF results for the 100 bootstrap trials.
Using this bootstrap approach, we find errors that are $50\% - 100\%$ larger than the errors from a simpler Monte Carlo approach.

The Monte Carlo approach assumes 
the noise of the co-added galaxy spectrum to be representative of the noise for all the individual lenslets spectra that make up the co-added spectrum.
Since the individual lenslets that make up a bin often come from exposures that were taken several days (or months) apart, it is very likely that there exist significant differences in the noise properties from one lenslet spectrum to the next (caused by variations in nighttime conditions, instrument performance, or spatially within the galaxy). 
This variation is captured only to a lesser extent by the noise estimates that are used in the Monte Carlo approach, since these are generated from the final spectrum after co-adding all the input lenslet spectra.
We believe that the errors from the bootstrap approach are instead more realistic since this approach doesn't make any assumptions about the noise properties of the data and incorporates any systematic differences between spectra extracted from different science frames and lenslets prior to co-adding.


\begin{figure*}
    \plottwo{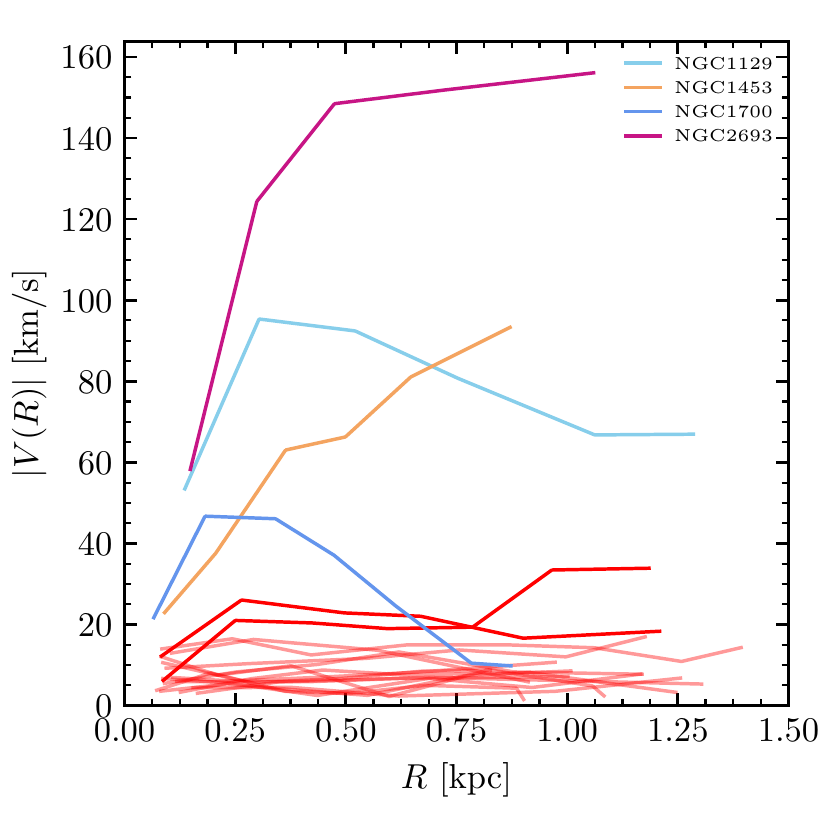}{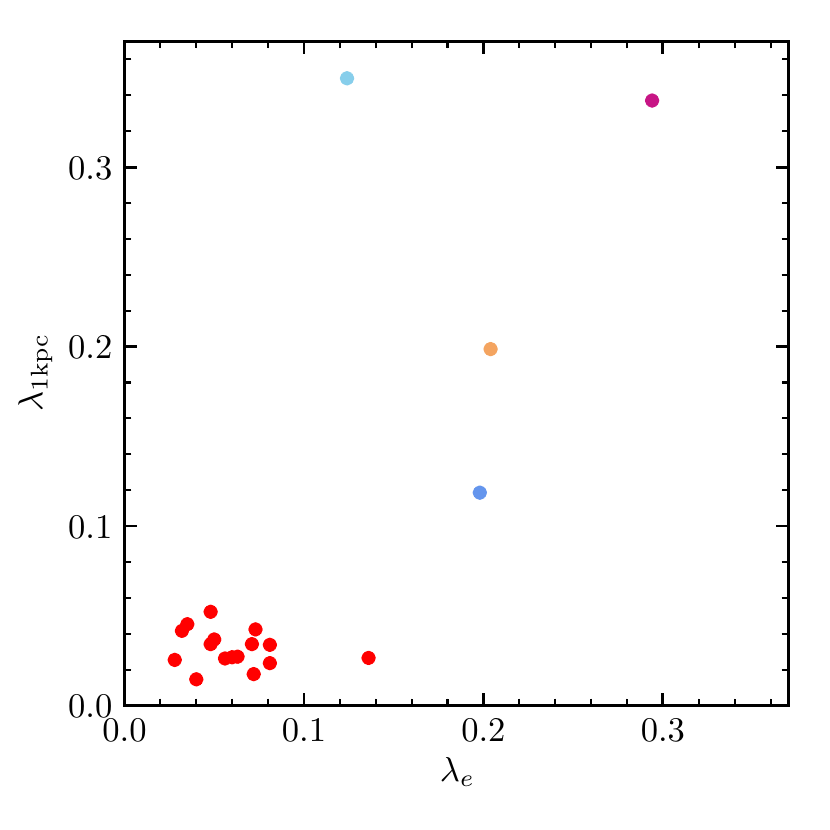}
    \caption{
        Left: Velocity profiles measured from the GMOS data along the photometric major axis of the 20 MASSIVE galaxies studied in this paper.
        Four galaxies -- NGC~1129, NGC~1453, NGC~1700 and NGC~2693 -- show prominent rotational features in the inner 1 to 1.5 kpc; the remaining 16 galaxies all have $|V| \lesssim 30$ km s$^{-1}$ (red curves).
       NGC~1700 (dark blue) has a counter-rotating core that is $180^\circ$ misaligned with the main-body rotation (see Fig.~\ref{fig:1700-vmap}).
        Right: Core spin parameter ($\lambda_{\rm 1 kpc}$) from GMOS IFS vs main-body spin parameter ($\lambda_e$) from Mitchell IFS for the 20 MASSIVE galaxies.  A circular aperture of radius 1 kpc and $1 R_e$ is used, respectively.
        The four galaxies with clear central rotations ($\lambda_{\rm 1 kpc} \gtrsim 0.1$) also have high global spins ($\lambda_e \gtrsim 0.1$), while the other galaxies all have low spins ($\lambda_{\rm 1 kpc}, \lambda_e \lesssim 0.1$).
    }
    \label{fig:v-profiles}
\end{figure*}

\section{Velocity features}
\label{sec:velocity_features}

Fig.~\ref{fig:v-profiles} (left panel) shows the radial velocity profiles measured from our GMOS IFS data along the photometric major axis of the central region of each of the 20 galaxies.
Sixteen of the 20 galaxies have low rotation velocities with $|V| \lesssim 30$ \kms.
Three galaxies exhibit fast central rotation, with $|V|$ reaching values of $\sim 100-150$ \kms: NGC~1129 (light blue), NGC~1453 (orange), and NGC~2693 (magenta). 
The rotation speeds at $R\sim 1$ to 1.5 kpc ($3''$ to $4''$) of these three galaxies in Fig.~\ref{fig:v-profiles} are maintained at a similar
level out to one effective radius and beyond in our wide-field Mitchell IFS data, which measured a maximum $|V|$ of 70 \kms, 95 \kms, and 150 \kms\ for NGC~1129, NGC~1453, and NGC~2693, respectively (Table 1, \citealt{Eneetal2018}).

To quantify the importance of rotation relative to dispersion in the central region probed by the GMOS IFS, we compute the spin parameter $\lambda$ measured within a circular aperture of radius 1 kpc, $\lambda_{\rm 1 kpc}$, where $\lambda(<R)\equiv \langle R|V|\rangle / \langle R\sqrt{V^2+\sigma^2}\rangle$, and the brackets refer to luminosity-weighted averages.  
Individual values of  $\lambda_{\rm 1 kpc}$ are listed in Table~\ref{tab:galaxy_props} (column 8) and plotted in the right panel of Fig.~\ref{fig:v-profiles}.
When compared to the larger-scale spin measured within one effective radius, $\lambda_e$ (column 6 of Table~\ref{tab:galaxy_props}),
Fig.~\ref{fig:v-profiles} (right panel) shows
that the galaxies with higher $|V|$ featured in the left panel also have larger $\lambda_{\rm 1 kpc}$ as well as larger $\lambda_e$.  The majority of our sample galaxies, however,
has both low central and low global spins and clusters in a distinct part of the $\lambda_{\rm 1 kpc}$-$\lambda_e$ parameter space in Fig.~\ref{fig:v-profiles}.

\begin{figure}
    \includegraphics[width=\columnwidth]{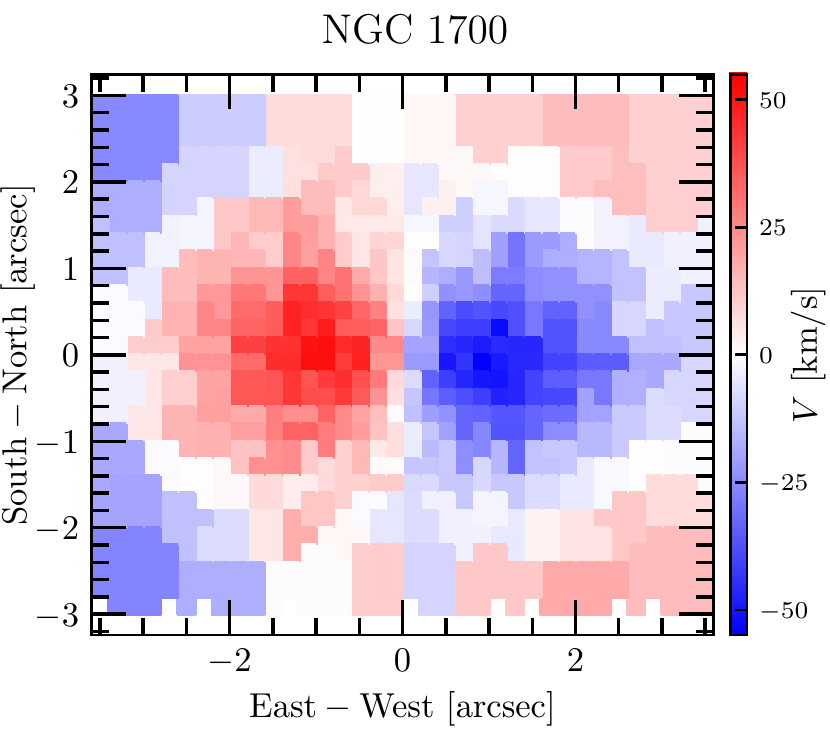}
    \caption{ Two-dimensional GMOS velocity map of the inner $\sim 2$ kpc by 2 kpc region of NGC~1700.  The central kinematically distinct component extends across most of the GMOS FOV and rotates in the opposite direction from the main body of the galaxy.
    }
    \label{fig:1700-vmap}
\end{figure}

The fourth galaxy highlighted in the left panel of Fig.~\ref{fig:v-profiles}, NGC~1700 (dark blue), shows a prominent KDC in our GMOS data. This component rotates in exactly the opposite direction of the main body rotation (see Fig.~\ref{fig:1700-vmap}). 
The existence of this counter rotating core (CRC) was first pointed out by \cite{Franxetal1989} and later examined through detailed stellar kinematics up to four effective radii of \cite{Statleretal1996}.
Both studies used long-slit spectroscopic observations and neither achieved the high angular resolutions and high $S/N$ presented here.

Previous works have also uncovered kinematically distinct features at the center of ETGs.
\citet{Riccietal2016} studied the stellar kinematics of the circumnuclear regions of 10 massive, nearby ETGs and found a KDC with an extent of $\sim 200$ pc,
while \cite{McDermidetal2006} found that some fast rotating SAURON ETGs have central small scale counter-rotating KDCs and that these KDCs have stellar populations that are younger than those of the main galaxy body.
This is also the case for the NGC~1700 KDC, which \cite{Kleinebergetal2011} found to have a distinct stellar population, much younger than the main body and suggest that it formed as the result of a merger between the main galaxy and a small stellar companion on a retrograde orbit.
On larger scales ($R \sim 10''$), two KDCs in MASSIVE galaxies NGC~507 and NGC~5322 were detected in our wide-field Mitchell IFS data \citep{Eneetal2018}. 
The rotation axis of the inner component of NGC~507 was misaligned by $\sim 105 \degr$ from the rotation axis of the outer component, while for NGC~5322 the two components are misaligned by almost $180 \degr$.
We noted in \citet{Eneetal2018} that more MASSIVE galaxies may contain KDCs on smaller scales not resolvable by the $4.1''$ diameter fibers of the Mitchell IFS.  NGC~1700 is one such example.

Kinematic features such as central KDCs and any misalignment between the kinematic and the photometric axes provide useful clues about the merger history of massive ellipticals.  Visual inspections of the 20 GMOS velocity maps in Appendix~\ref{sec:kinematic_moments_plots} suggest that the kinematic axis of the central $\sim 8''$ region -- when rotation is detected -- is not necessarily aligned with the photometric axis (which is typically the long axis of the GMOS FOV).  The relative angle ranges from well aligned (e.g., the 3 galaxies with high core rotations) to almost maximally misaligned (e.g., NGC~1700).  In addition, the GMOS data now enable us to quantify how the kinematic features in the central regions of massive galaxies are connected to and aligned with the large-scale kinematic features measured from our Mitchell data in \citet{Eneetal2018}. A detailed analysis of these properties will be presented in a separate work.


\section{Velocity dispersion}
\label{sec:velocity_dispersion}

\begin{figure}
    \includegraphics[width=\columnwidth, page=1]{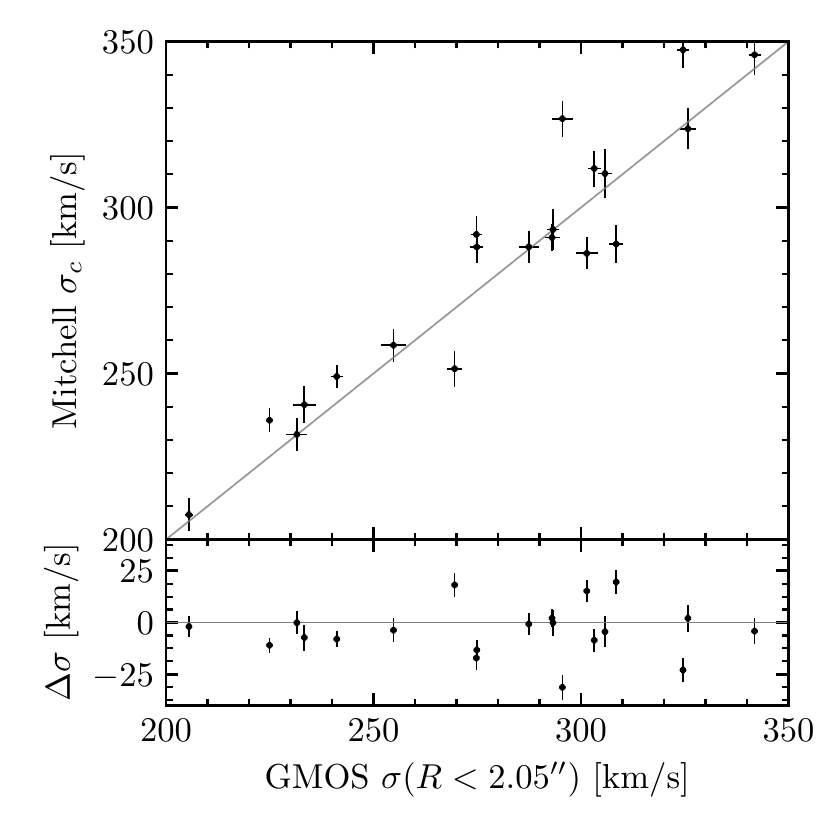}
    \caption{Comparison of central stellar velocity dispersion $\sigma$ measured from GMOS vs Mitchell over the same aperture for the 20 MASSIVE galaxies (top).
        The GMOS $\sigma$ is computed as a luminosity-weighted average within a circular aperture of $4.1''$ diameter, the size of a Mitchell fiber.  The Mitchell $\sigma$ is measured from the central Mitchell fiber.
        The one-to-one line is shown in light gray.
        The difference between the two velocity dispersions is shown in the bottom panel.  The error on the difference is the quadrature sum of the GMOS and Mitchell errors.
        The overall agreement between the GMOS and Mitchell measurements is excellent, with a median fractional differene of $({\rm GMOS}~\sigma - \sigma_{c})/\sigma_{c} = -0.01 \pm 0.02$.
    }
    \label{fig:sigma_gmos_versus_vp}
\end{figure}

\subsection{Central velocity dispersion}
\label{sub:central_velocity_dispersion}

\begin{figure*}
    \includegraphics[width=\textwidth]{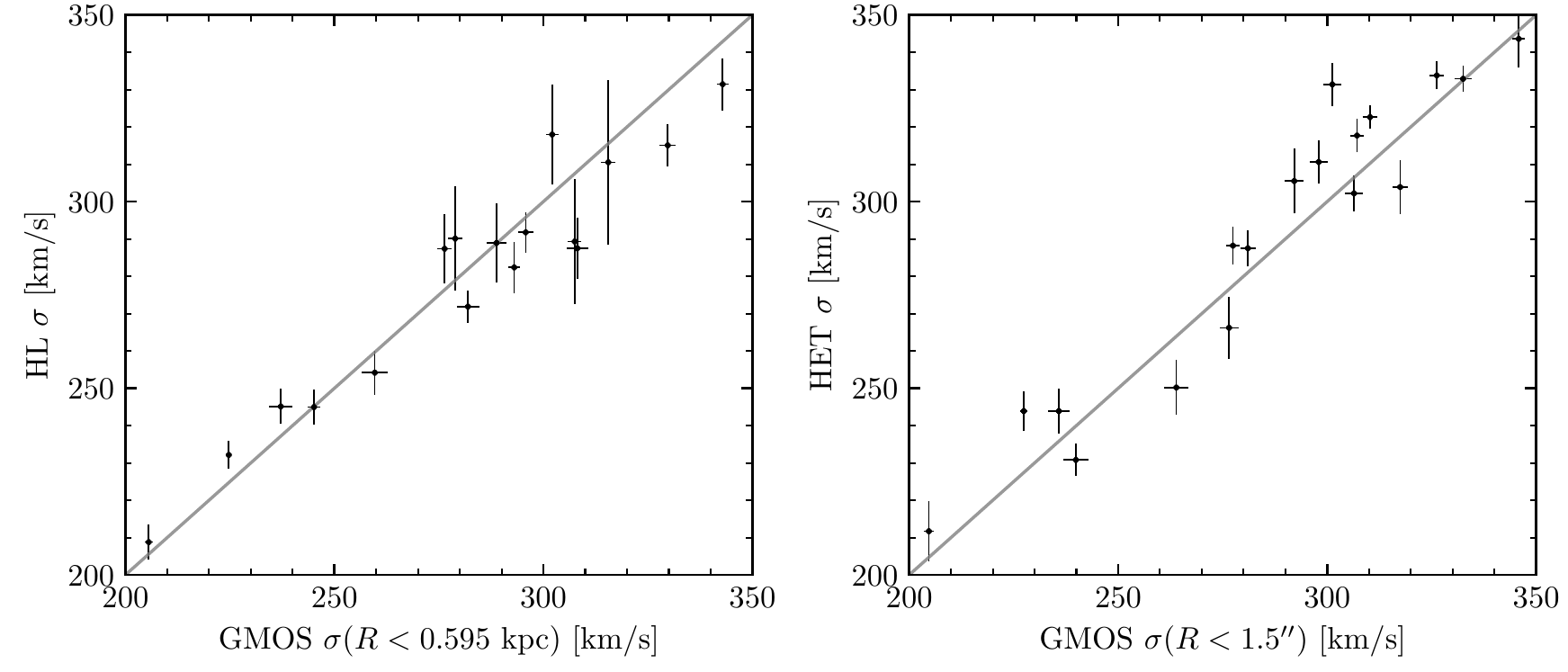}
    \caption{Comparison of the GMOS stellar velocity dispersions  to literature values from the HyperLeda catalog (left; \citealt{Patureletal2003}) and the HET catalog (right; \citealt{vandenBoschetal2015}). 
        For a fair comparison, the GMOS value in each panel is computed as a luminosity-weighted average within a circular aperture comparable to that used in the catalog (a radius of 0.595 kpc for HyperLeda and a radius of $1.5''$ for HET).
        The light gray line indicates the one-to-one relation.  
        The median fractional difference is $({\rm GMOS}~\sigma - {\rm HL}~\sigma)/{\rm HL}~\sigma = 0.01 \pm 0.02 $ for Hyperleda and  $({\rm GMOS}~\sigma - {\rm HET}~\sigma)/{\rm HET}~\sigma = -0.02 \pm 0.02 $ for HET.
    }
    \label{fig:sigma_lit}
\end{figure*}

The high spatial sampling of the GMOS IFS enables us to measure the stellar velocity dispersion $\sigma$ within apertures of different sizes. 
We measure the central stellar velocity dispersion by taking a luminosity-weighted average of $\sigma$ within a circular aperture of radius 1 kpc. 
The resulting values are listed in column 9 of Table~\ref{tab:galaxy_props}.
We find that 6 of the 20 galaxies have high central velocity dispersion $\sigma_{\rm 1 kpc} \gtrsim 300$ \kms, 4 have $\sigma_{\rm 1 kpc} \lesssim 230$ \kms, and the rest is in between. 
 NGC~890 has the lowest value with $\sigma_{\rm 1 kpc}=206$ \kms.

As another measure of central $\sigma$, we take a luminosity-weighted average of the GMOS $\sigma$ within an aperture of $4.1''$ diameter, which is the size of one Mitchell IFS fiber.  This GMOS $\sigma$ can then be compared directly to the Mitchell $\sigma$ 
measured from the central Mitchell fiber. Fig.~\ref{fig:sigma_gmos_versus_vp} shows an excellent overall agreement, with a median fractional difference of $({\rm GMOS}~\sigma - \sigma_{c})/\sigma_{c} = -0.01 \pm 0.02$.

The observed rms scatter in the difference of the two velocity dispersion measurements is $\sim 12$ \kms, while the estimated scatter computed from the individual measurement errors is $\sim 6$ \kms.
Since the GMOS $\sigma$ is an average of the velocity dispersion over many spatial bins, each of which is measured from a high S/N spectrum, the statistical errors on the GMOS $\sigma$ in Fig.~\ref{fig:sigma_gmos_versus_vp} are very small ($\sim 1-2$ \kms) and are likely subdominant to various systematic errors.
Using the measurement errors on individual GMOS bins ($\sim 10$ \kms) instead of the luminosity-averaged statistical erros, we find that the scatter in the difference of the velocity dispersions is in good agreement with the scatter estimated from the errors.
Keeping this in mind and considering that the GMOS and Mitchell $\sigma$ are measured using data from different instruments and telescopes over different spectral regions using different template libraries, the agreement shown in Fig.~\ref{fig:sigma_gmos_versus_vp} is nonetheless reassuring.
A similar comparison for the higher-order moments $h_3$ and $h_4$ finds that the observed rsm scatter is $\sim 0.025$ while the scatter estimated from the measurement errors is $\sim 0.015$. 
Once again, accounting for the fact that the luminosity-averaged statistical errors are underestimating the observed errors ($\sim 0.02$), we recover a much better agreement between the observed and estimated scatter.

We note that measuring the GMOS $\sigma$ as a luminosity-weighted average over the dispersions from individual bins within some spatial region is not identical to measuring $\sigma$ from a single co-added spectrum within the same spatial region (e.g., a Mitchell fiber sized aperture).
To quantify this difference, we use the GMOS data to generate a co-added spectrum (where we sum all the individual lenslet spectra that fall within a $4.1''$ diameter aperture) and derive the pPXF best-fitting kinematic moments for the co-added spectrum.
Although the $\sigma$ values computed through these two methods (luminosity-weighted average $\sigma$ and $\sigma$ from a co-added spectrum) are not identical, we found the differences to be small, on the order of $\lesssim 5\%$.
A few outliers correspond to galaxies with large velocity gradients in the region probed by GMOS, which the $4.1''$ diameter Mitchell fibers do not resolve.
The unresolved high velocity leads to an increase in the recovered value of Mitchell velocity dispersion.
If we instead compute $v_{rms} = \sqrt{V^2 + \sigma^2}$, we find better agreement with the Mitchell value.
\citet{Vealeetal2017a} also found agreement between $\sigma$ computed from these two methods.

\subsection{Comparison with literature}
\label{sub:comparison_with_literature}

To compare the GMOS measurements with literature values of $\sigma$ compiled in the HyperLeda catalog \citep{Patureletal2003},
we compute the GMOS $\sigma$ as a luminosity-weighted average within the aperture used by HyperLeda (0.595 kpc radius).  The left panel of Fig.~\ref{fig:sigma_lit}
shows good agreement between GMOS and HyperLeda values, with a median fractional difference of $({\rm GMOS}~\sigma - {\rm HL}~\sigma)/{\rm HL}~\sigma = 0.01 \pm 0.02 $.  The HyperLeda $\sigma$ values are compiled from heterogeneous measurements in the literature and usually have large error bars.

Deviations of more than $10\%$ occur for three galaxies: NGC~315 ($325 \pm 1$ \kms~versus $296 \pm 10$ \kms~in HyperLeda), NGC~1129 ($237 \pm 3$ \kms~versus $326 \pm 15$ \kms~in HyperLeda), and NGC~2693 ($299 \pm 3$ \kms~versus $339 \pm 10$ \kms~in HyperLeda).
We note that the HyperLeda catalogue contains a large range of values for NGC~315 (260 \kms~to 360 \kms) and NGC~2693 (270 \kms~to 400 \kms).

For 18 galaxies in our sample, we can also compare the GMOS $\sigma$ with those reported in the HET catalogue \citep{vandenBoschetal2015}.  
The HET catalogue provides stellar kinematic parameters for 1022 galaxies measured from long-slit spectra in the wavelength range 4200 -- 7400 \AA~taken with the Marcario Low Resolution Spectrograph \citep{Hilletal2008c}
on the Hobby-Eberly Telescope (HET) at McDonald Observatory.
The HET values are measured using a $3.5'' \times 2''$ aperture and the typical S/N inside the aperture is greater than 100.
The right panel of Fig.~\ref{fig:sigma_lit}
compares the HET $\sigma$ with our GMOS $\sigma$ measured within an equivalent circular aperture with radius $1.5''$. The median fractional difference is $({\rm GMOS}~\sigma - {\rm HET}~\sigma)/{\rm HET}~\sigma = -0.02 \pm 0.02$.
We also note that for the 3 outliers with HyperLeda mentioned in the previous paragraph, our values agree better with HET:  $334 \pm 4$ \kms~for NGC~315, $231 \pm 4$ \kms~for NGC~1129, and $331 \pm 6$ \kms~for NGC~2693.

\subsection{Aperture correction}
\label{sub:aperture_correction}

Aperture correction relations are used frequently to transform velocity dispersions measured with fiber-fed spectrographs such as SDSS to values measured within a uniform aperture of a standard physical size.
These correction relations are important for systematic studies of galaxies at different distances and particularly so in the case of high redshift studies where the typical IFS fiber size covers a significant fraction of the small apparent size of the galaxies.
A typical application consists of correcting the velocity dispersion of galaxies to apertures with sizes related to the effective radius, e.g., $R_e$, $R_e/2$, or $R_e/8$.
Our GMOS and Mitchell IFS data together span a wide radial coverage (from $\sim 0.5''$ to beyond $50''$) of each galaxy
and can be used to derive an aperture correction relation between velocity dispersions measured within two commonly used radii, $R_e/8$ and $R_e$.

For most of our galaxies, the $5'' \times 7''$ coverage of the GMOS FOV corresponds to a radial extent of $\sim 1-1.5$ kpc from the center of each galaxy, allowing us to measure the value of $\sigma$ within $R_e/8$.  Here we use $R_e$ measured from deep $K$-band photometric data taken with WIRCam on the Canada-France-Hawaii Telescope as part of the MASSIVE Survey (Quenneville et al. in prep).  To determine $R_e$, 
the photometry package ARCHANGEL \citep{Schombert2007} is used to fit elliptical isophotes to the stacked image of each galaxy.  A curve of growth is then constructed from the aperture luminosity for each isophote as a function of radius. The total luminosity and half-light radius $R_e$ are then measured from the curve of growth.  
The values of $R_e$ for our sample of 20 galaxies are given in column 5 of Table~\ref{tab:galaxy_props}.  They range from $16''$ to $45''$, with the average $R_e$ being $25''$ ($\sim 9$ kpc).

\begin{figure}
    \includegraphics[width=\columnwidth]{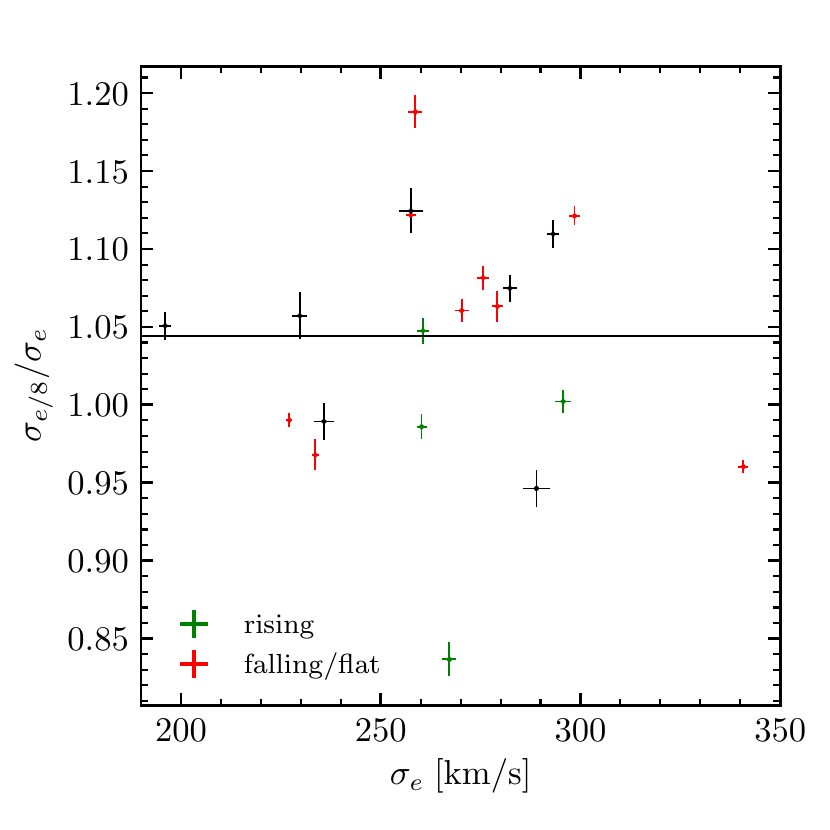}
    \caption{Aperture correction for velocity dispersions measured within $R_e/8$ and $R_e$. 
        The dispersions within $R_e/8$ are measured using GMOS data.
        The dispersions at $R_e$ are measured using Mitchell data.
        Colors denote the behavior of the outer velocity dispersion profile as quantified by \citet{Vealeetal2018}: rising (green), falling/flat (red) or not quantified (black).
        The solid black line shows the average ratio $\sigma_{e/8}/\sigma_e = 1.044 \pm 0.078$. 
        Previous studies have found $\sigma_{e/8}/\sigma_e = 1.062 \pm 0.079$ \citep{Vealeetal2017a}, $1.147\pm 0.083$ (\citealt{cappellarietal2006}),
        1.087 (\citealt{Jorgensenetal1995}), and 1.133 (\citealt{Mehlertetal2003}).
    }
    \label{fig:aperture_correction}
\end{figure}

In Fig.~\ref{fig:aperture_correction} we present the aperture correction relation between the velocity dispersion at $R_e/8$ (measured from GMOS data) and the velocity dispersion at $R_e$ (measured from Mitchell data).
We find the mean correction to be only 4.4\% but the scatter is large:
\begin{equation}
\frac{\sigma_{e/8}}{\sigma_e} = 1.044 \pm 0.078.
\label{eq:ratio}
\end{equation}
The fact that the average of the ratio of $\sigma$ at the two radii is close to unity by no means implies that the radial profiles of $\sigma$ are flat between $R_e/8$ and $R_e$.  The diverse radial profiles of $\sigma$ are evident in Figs.~\ref{fig:A1}-\ref{fig:A10} of Appendix~\ref{sec:kinematic_moments_plots} and are reflected in the the large scatter in Eqn.~\ref{eq:ratio}.

Four galaxies in our sample were identified in \cite{Vealeetal2018} as having rising outer velocity dispersion profiles (green points in Fig.~\ref{fig:aperture_correction}).  These galaxies have higher $\sigma$ at $R_e$ than at $R_e/8$. When these four galaxies are excluded, the ratio increases to $1.062 \pm 0.076$.

The fact that the ratio of the two $\sigma$ measurements is so close to unity could also be caused by luminosity-weighting, which biases $\sigma_e$ to higher values by assigning more weight to the inner bins.
On average, the luminosity-weighted $\sigma_e$ is $\sim 5-10$ \kms~ higher than the corresponding arithmetic average.
Using the arithmetic average value for $\sigma_e$ for all 20 galaxies leads to a ratio of $1.064 \pm 0.084$.

\begin{figure*}
    \includegraphics[width=\textwidth]{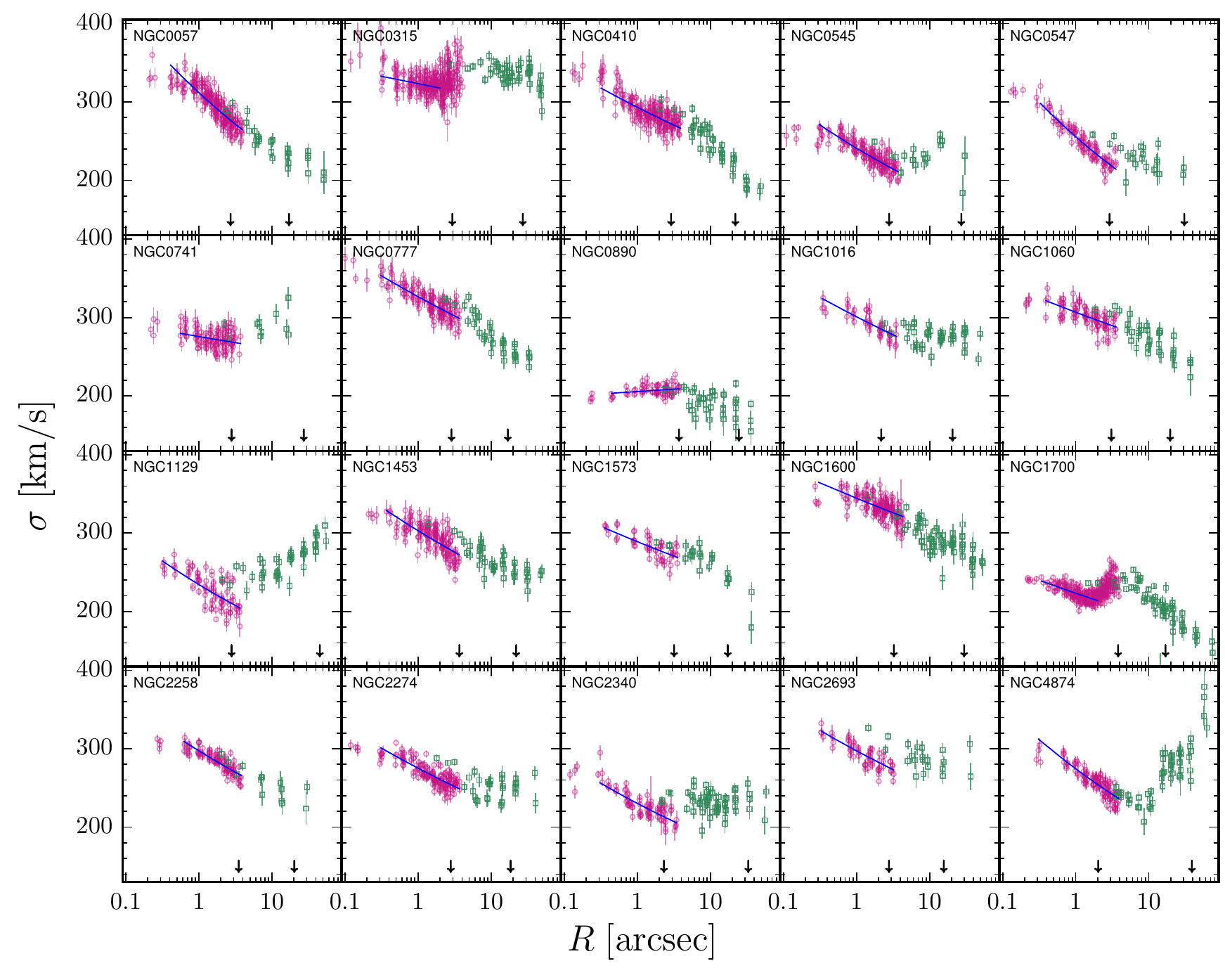}
    \caption{Radial profiles of the stellar velocity dispersion measured from the GMOS (purple circles) and Mitchell (green squares) IFS for the 20 MASSIVE galaxies in this study.
    The two IFS datasets together cover two orders of magnitude in radius that is plotted on a logarithmic scale.
    At a given radius, the multiple data points represent $\sigma$ measured in the various angular locations of the IFS.
    The GMOS data provide finely-resolved stellar kinematics within about 1 kpc (indicated by the inner arrow) of each galaxy, while the Mitchell data provide wide-field kinematic measurements to one effective radius (outer arrow) or beyond.
    The two datasets agree well in the overlapping region at $R \sim 3''-4''$.
    The best-fit power law form to each GMOS $\sigma(R)$ is over-plotted (blue lines), showing that 19 of the 20 galaxies have $\sigma(R)$ that rise towards the center.
    Notable exceptions to a single power law fit to $\sigma(R)$ across the GMOS FOV are NGC~315 and NGC~1700.
    }
    \label{fig:sigma_profiles}
\end{figure*}

The aperture correction reported here is in agreement with our earlier measurement of $\sigma_{e/8}/{\sigma_e} = 1.062 \pm 0.079$ using the Mitchell data for the 41 brightest MASSIVE galaxies \citep{Vealeetal2017a}.
For 40 ETGs with lower mass than our sample,
the SAURON IFU data gave a higher mean aperture correction but a comparably large scatter: $1.147 \pm 0.083$ \citep{cappellarietal2006}.
Other aperture correction studies for nearby ETGs
found a ratio of 1.087 using a compilation of kinematic and photometric data from literature for 51 E and S0 galaxies \citep{Jorgensenetal1995}, and a ratio of 1.133
from long-slit spectroscopic data for 35 ETGs in the Coma cluster \citep{Mehlertetal2003}, but 
neither work reported error bars.

\subsection{Velocity dispersion radial profiles}
\label{sub:velocity_dispersion_radial_profiles}

\begin{figure}
    \includegraphics[width=\columnwidth]{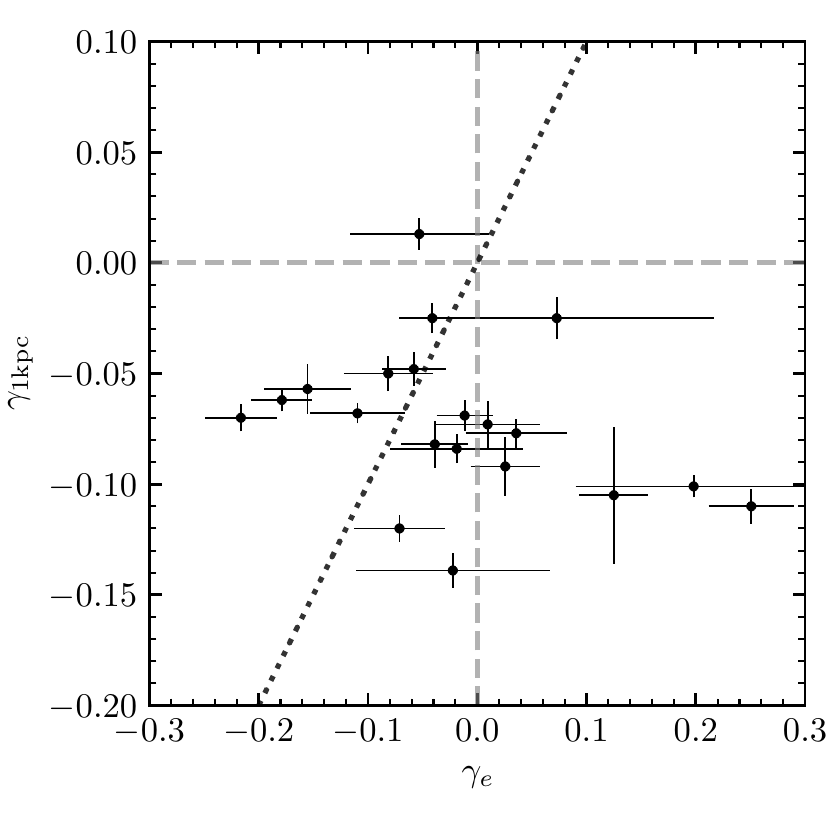}
    \caption{Logarithmic slope of the radial profile of stellar velocity dispersion in the central 1 kpc ($\gamma_{\rm 1 kpc}$) versus at one effective radius ($\gamma_{e}$) for the 20 MASSIVE galaxies in this study. 
    While 19 out of the 20 galaxies have negative $\gamma_{\rm 1 kpc}$ (see also Fig.~\ref{fig:sigma_profiles}), the outer slope $\gamma_{e}$ spreads from negative to positive values.
    Some galaxies are consistent with a single power law ($\gamma_{\rm 1 kpc}=\gamma_{e}$, dotted line) over the combined radial range of the GMOS and Mitchell data. 
    }
    \label{fig:gammain-gammaout}
\end{figure}

Fig.~\ref{fig:sigma_profiles} shows the velocity dispersion $\sigma$ as a function of radius from $\sim 0.3''$ to $\sim 50''$ from our GMOS (purple circles) and Mitchell (green squares) IFS data.
For more details, see the individual kinematic maps and radial profiles (unfolded along the IFU PA) of the four velocity moments $V, \sigma, h_3$, and $h_4$ in
Appendix~\ref{sec:kinematic_moments_plots}.
In the inner few arcsecs, the $\sigma$ profiles
for most galaxies in our sample decrease from peak values close to the galaxy centers to smaller values further out.  

In the transition region at $3''$ -- $4''$, the GMOS values are well matched to the Mitchell values, with the exception of a few cases such as NGC~1129 and NGC~2693.
These galaxies represent the outliers mentioned at the end of Section~\ref{sub:central_velocity_dispersion} with high GMOS velocity gradients that cannot be resolved by the $4.1''$ diameter Mitchell fiber(s) and for which the velocity dispersion of the inner few Mitchell fibers should be compared to $v_{rms} = \sqrt{V^2 + \sigma^2}$ measured from the GMOS data.
It should also be noted that the exact location
 of the innermost Mitchell data point in each radial plot in Fig.~\ref{fig:sigma_profiles} is uncertain to within a few arcsecs due to the large
diameter of the Mitchell fiber.

For most galaxies $\sigma$ continues to decrease in the region probed by Mitchell data.
We find that $\sigma$ at large radii is typically smaller than near the center by $\sim 100$ -- 150 \kms.
Four galaxies -- NGC~545, NGC~741, NGC~1129 and NGC~4874 -- have rising $\sigma$ profiles towards larger radii, with NGC~1129 and NGC~4874 showing the most prominent rise.
The one outlier
is NGC~890, whose $\sigma$ has a peak value at $R \sim 3''$ and decreases gently both towards smaller and larger radii.

We quantify $\sigma(R)$ in the central region probed by GMOS using a single power law: $\sigma(R) \propto R^{\gamma_{\rm 1 kpc}}$.
The best-fit logarithmic slope $\gamma_{\rm 1 kpc}$ for each galaxy is given in column 10 of Table~\ref{tab:galaxy_props}, and
each best-fitting profile is shown as a blue straight line over the radial range used in the fit in Fig.~\ref{fig:sigma_profiles}.
We use only data points beyond $R \approx 0.3''$ for the fit to mitigate the effects of seeing that can flatten the shape of $\sigma(R)$ in the innermost part.
The $\sigma$ profiles from $0.3''$ to $\sim 4''$ are reasonably approximated by a single power law,
with the exception of NGC~315 and NGC~1700, which show a break at $R \sim 2''$.
For these two galaxies we restrict the fit to $R < 2''$.
Our aim here is to find a simple form to approximate the shape of the GMOS $\sigma(R)$ outside of $R \sim 0.3''$, and for this purpose, a single power-law provides a reasonable fit to the data.
All galaxies have $\chi^2$ per degree of freedom (DOF) $\sim 3$ or less, with the exception of NGC~1129 which has $\chi^2$ per DOF $\sim 7$.
We do not think it is worthwhile to attempt to find a closer fit to the diverse shapes of $\sigma(R)$ using a more complicated functional form with PSF convolution. Instead, we will perform full dynamical modeling using the complete 2D measurements of $\sigma$ as well as other kinematic moments and will report the results in later papers.

We find that 19 out of the 20 galaxies have negative $\gamma_{\rm 1 kpc}$, i.e., $\sigma$ increases from $\sim 3''$ inward.  
To compare the central behavior of $\sigma(R)$ with that at larger radii, we compare $\gamma_{\rm 1 kpc}$ with the logarithmic slope of $\sigma(R)$ at the effective radius.  
For the latter, we measure $\gamma_{e}$ using the wide-field Mitchell data to fit a broken power law to $\sigma(R)$ \citep{Vealeetal2018}, and then use the asymptotic logarithmic slopes $\gamma_1$ and $\gamma_2$ to compute the \textit{local} logarithmic slope at $R_e$.
The two slopes $\gamma_{\rm 1 kpc}$ and $\gamma_{e}$ are plotted in Fig.~\ref{fig:gammain-gammaout}.  
In contrast to the negative $\gamma_{\rm 1 kpc}$, $\gamma_{e}$ is spread out from $-0.25$ to $+0.25$. 
Furthermore, more than half of the galaxies have $\gamma_{\rm 1 kpc} < \gamma_{e}$, i.e., $\sigma(R)$ changes logarithmic slopes in the inner $\sim 3''$ and rises more steeply towards the galaxy's center. 

\cite{Loubseretal2018} measured the slopes of the velocity dispersion profiles using long-slit spectroscopic observations of a combined sample of brightest cluster galaxies (BCGs) and brightest group galaxies (BGGs).
The radial coverage of their long-slit data extends up to $10$ kpc for BGGs and $15$ kpc for BCGs.
For the five GMOS galaxies in common with their BGG sub-sample, \cite{Loubseretal2018} found negative slopes.
The values are shown in Table~\ref{tab:sigma_slopes}, along with our own measurements of $\gamma_{\rm 1 kpc}$ and $\gamma_{e}$.
The \cite{Loubseretal2018} values of the velocity dispersion slope are in good agreement with our own $\gamma_{\rm 1 kpc}$ for all five galaxies and in moderate agreement with $\gamma_{e}$, except for NGC~410.

\begin{deluxetable}{cccc}
\tablecaption{Power law slopes of velocity dispersion profiles for 5 MASSIVE galaxies. \label{tab:sigma_slopes}}
\tablehead{
\colhead{Galaxy} & \colhead{$\gamma_{\rm 1 kpc}$} & \colhead{$\gamma_{e}$} & \colhead{$\gamma_{\rm Loubser}$}
}
\colnumbers
\startdata
NGC0315 & $-0.025 \pm 0.007$ & $-0.041 \pm 0.023$ & $-0.028 \pm 0.011$\\
NGC0410 & $-0.070 \pm 0.006$ & $-0.216 \pm 0.033$ & $-0.068 \pm 0.009$\\
NGC0777 & $-0.068 \pm 0.004$ & $-0.110 \pm 0.044$ & $-0.078 \pm 0.010$\\
NGC1060 & $-0.050 \pm 0.008$ & $-0.081 \pm 0.041$ & $-0.067 \pm 0.010$\\
NGC1453 & $-0.082 \pm 0.011$ & $-0.039 \pm 0.031$ & $-0.068 \pm 0.011$
\enddata
\tablecomments{
    (1) Galaxy name.
    (2) Velocity dispersion logarithmic slope at 1 kpc, measured from the GMOS IFS data.
    (3) Velocity dispersion logarithmic slope at the effective radius, measured from the Mitchell IFS data.
    (4) Velocity dispersion logarithmic slope measured by \cite{Loubseretal2018}.
    }
\end{deluxetable}

An increase in the line-of-sight $\sigma$ towards small radii can be accounted for by the presence of either a central mass concentration (e.g., a black hole) or radial anisotropy at small radii  \citep{BinneyMamon1982, Gerhardetal1998b}.
This mass-velocity anisotropy degeneracy can be broken somewhat by using information about the LOSVD kurtosis $h_4$, which we will discuss in Section ~\ref{sec:higher_order_moments}.
Our dynamical modeling of the full set of kinematic moments ($V, \sigma, h_3, h_4$) will determine which combination of velocity anisotropy and mass profiles would best fit the data presented in this paper.  
Prior such modeling of a handful of massive elliptical galaxies have found tangential velocity anisotropy and massive black holes in the central regions and radial velocity anisotropy at larger radii 
(\citealt{gebhardtetal2000,Gebhardtetal2003,ShenGebhardt2010,gebhardtetal2011,mcconnelletal2012,thomasetal2014,Thomasetal2016}).

\section{Higher-Order Velocity Moments}
\label{sec:higher_order_moments}

\subsection{Skewness $h_3$}

\begin{figure}
    \includegraphics[width=\columnwidth]{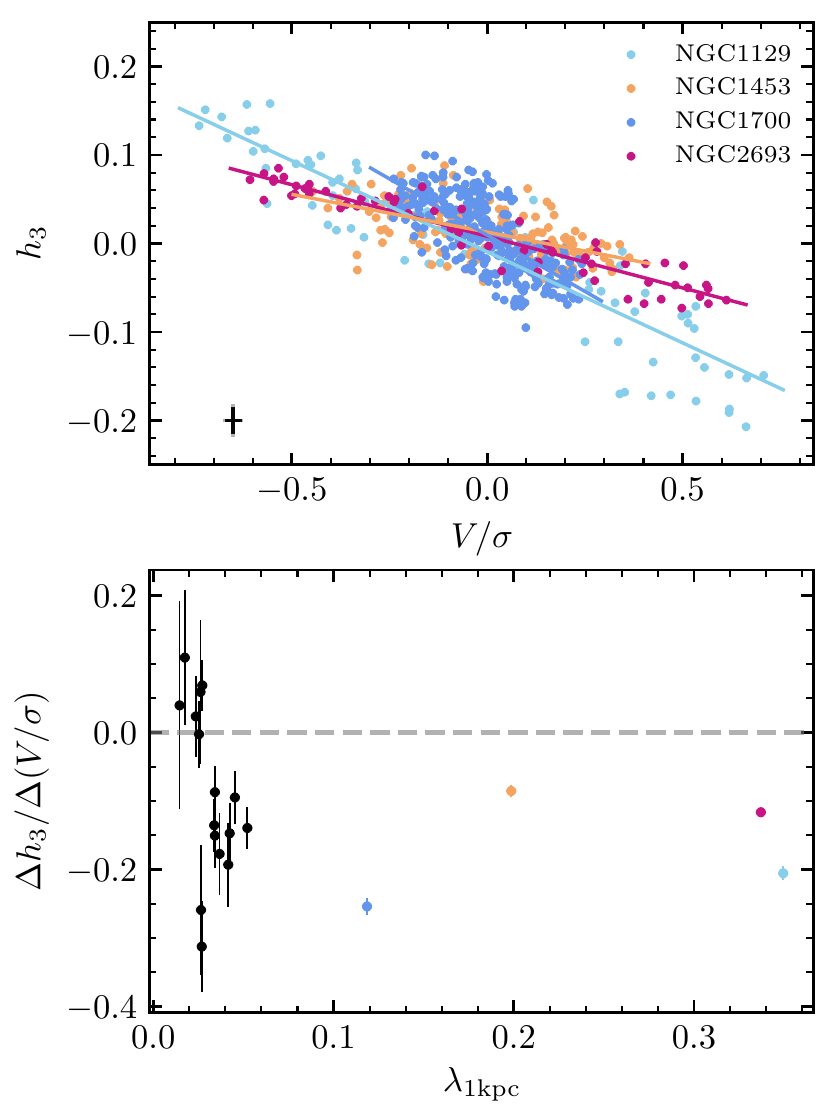}
    \caption{
        The two odd velocity moments, $V/\sigma$ and $h_3$, 
        are anti-correlated in the central region 
        of each of the 4 fast-rotating galaxies in our sample (top panel). 
        The solid lines show the best-fit anti-correlation slope. Typical error bars are shown in the lower left corner.
        In the bottom panel, the slope of the anti-correlation between $h_3$ and $V/\sigma$ is plotted against the central spin parameter $\lambda_{\rm 1 kpc}$. 
        High-spin galaxies ($\lambda_{\rm 1 kpc} \gtrsim 0.1$) exhibit clear anti-correlation with slopes of $\sim -0.1$ to $-0.25$.
        Galaxies with low spins ($\lambda_{\rm 1 kpc} \lesssim 0.1$) show a variety of slopes, some with large error bars due to the small range in $V/\sigma$.
    }
    \label{fig:h3-Vsig}
\end{figure}

In Fig.~10 of \cite{Vealeetal2017a}, we examined the behavior of the two odd velocity moments $V$ and $h_3$ 
on scales of $\sim 1$ kpc up to $\sim 20$ kpc measured from the Mitchell IFS data for a sample of MASSIVE galaxies. 
For the fast rotators ($\lambda_e \gtrsim 0.2$),
we found a clear anti-correlation between the spatially-resolved $h_3$ and $V/\sigma$ within each galaxy. The slope of the anti-correlation, $\Delta h_3/\Delta(V/\sigma)$, ranged between $-0.1$ and $-0.2$.  The slow rotators ($\lambda_e \lesssim 0.2$), on the other hand, showed positive, negative or no correlations between the two odd moments. 
This result is consistent with the interpretation that anti-correlations between $h_3$ and $V/\sigma$ within a galaxy are associated with its internal disc kinematics \citep{benderetal1994}, and only fast-rotating ETGs exhibit such features.

The high-resolution GMOS IFS data now enable us to extend this analysis to the core regions of 20 ETGs in the MASSIVE survey.
As shown in Fig.~\ref{fig:h3-Vsig} (top panel), we find a clear anti-correlation between $h_3$ and $V/\sigma$ for the four galaxies (NGC~1129, NGC~1453, NGC~1700, NGC~2693) with noticeable core rotations in Fig.~\ref{fig:v-profiles}. 
Since the rotations continue from $\sim 1$ kpc to $1R_e$ and beyond (see Section~\ref{sec:velocity_features}), these four galaxies have relatively high spin parameters (measured at both 1 kpc and $1 R_e$): {\bf $\lambda_{\rm 1 kpc} = 0.350, 0.199, 0.119, 0.337$} and $\lambda_e = 0.124, 0.204, 0.198$ and $0.294$, respectively. 
 
The slope of the spatially-resolved $h_3$ and $V/\sigma$ measured within each galaxy is plotted against the galaxy spin parameter for all 20 MASSIVE galaxies in the bottom panel of Fig.~\ref{fig:h3-Vsig}.
For the faster rotators with $\lambda_{\rm 1 kpc}$ above 0.1, the anti-correlation slopes are between $-0.1$ and $-0.25$, similar to the range measured on larger scales in
\cite{Vealeetal2017a}.
The large spread in the slopes with both positive and negative values for the remaining galaxies is also similar to
that seen on large scales for slow rotators in
\cite{Vealeetal2017a}.

As described in Section~\ref{sec:velocity_features}, NGC~1700 has a counter-rotating core that rotates in exactly the opposite direction from the main body rotation.
Despite its relatively low central rotation ($\sim 40$ \kms; Fig.~3), the $h_3$ and $V/\sigma$ within the core region of NGC~1700 show a steep anti-correlation with a slope of $-0.25$.
Hints of $h_3 - V/\sigma$ anti-correlation for galaxies with counter-rotating cores or kinematically decoupled components have also been found in lower-mass ETGs in the ATLAS$^{\rm 3D}$ survey \citep{krajnovicetal2011}.

\subsection{Kurtosis $h_4$}

\begin{figure}
    \includegraphics[width=\columnwidth]{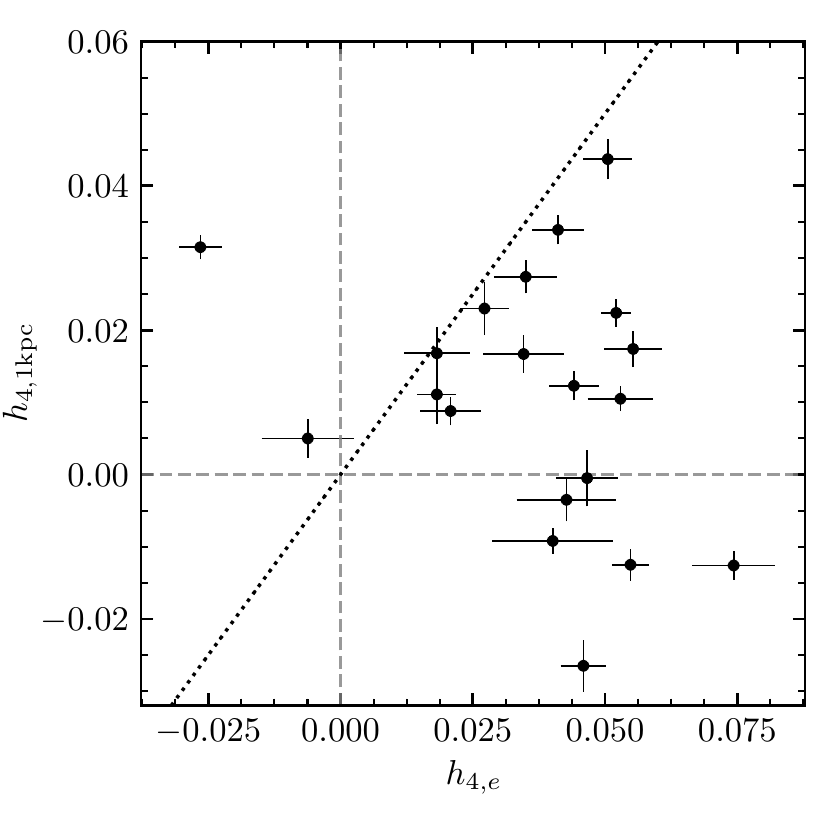}
    \caption{
        Luminosity-averaged kurtosis $h_4$ measured within $1$ kpc using GMOS data versus luminosity-averaged $h_4$ within one effective radius measured from Mitchell data.
        The dotted line indicates the one-to-one relation.
        On both small and large scales, the average $h_4$ is predominantly positive.    }
    \label{fig:avg_h4_gmos_vp}
\end{figure}

In Fig.~\ref{fig:avg_h4_gmos_vp} we compare the luminosity-weighted average $h_4$ measured within $1$ kpc using the GMOS data to the luminosity-weighted average $h_4$ measured within one effective radius using the Mitchell data.
On both small and large scales, we find that the average $h_4$ is predominantly positive: 14/20 galaxies have $h_{4, 1 kpc} > 0$ and 18/20 galaxies have $h_{4, e} > 0$.
We have previously found that MASSIVE galaxies show predominantly positive large scale average $h_4$ that is in $\sim 50 \%$ of cases accompanied by a rising outer velocity dispersion profile (\citealt{Vealeetal2017a, Vealeetal2018}).

As mentioned in Section~\ref{sec:velocity_dispersion}, $h_4$ can help break the degeneracy between the effects of mass and velocity anisotropy on the behavior of the line-of-sight velocity dispersion.
For example, \cite{Vealeetal2018} found a statistically significant positive correlation between the radial gradient in $h_4$ and the velocity dispersion outer logarithmic slope $\gamma_{\rm outer}$ for a sample of 90 MASSIVE galaxies.
They argue that this trend is more likely due to mass profile variations than to velocity anisotropy, since radial anisotropy at large radii would imply that a positive $h_4$ gradient comes along with a more negative $\sigma$ gradient.
Their arguments applied to the behavior of $\sigma$ and $h_4$ (and the implications for mass and velocity anisotropy) at large radii, where the dominant contribution comes from the dark matter halo mass.

\begin{figure}
    \includegraphics[width=\columnwidth]{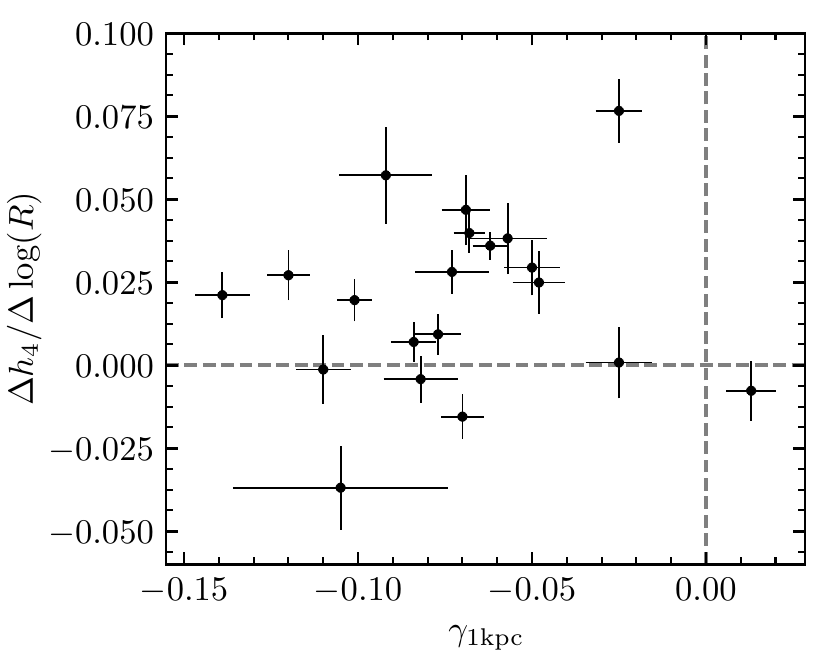}
    \caption{
        Radial gradient of the kurtosis $h_4$ versus radial gradient of the velocity dispersion $\gamma_{\rm 1 kpc}$, both measured in the inner $\sim 1$ kpc of the 20 MASSIVE galaxies.
        The majority of the galaxies has rising $\sigma(R)$ and falling $h_4(R)$ towards the center.}
    \label{fig:h4_mk_rgrad}
\end{figure}

By contrast, the GMOS data studied here probe the central regions of galaxies in which stars and possibly black holes dominate the mass.  
In Fig.~\ref{fig:h4_mk_rgrad} we plot the $h_4$ radial gradient versus the velocity dispersion logarithmic slope $\gamma_{\rm 1 kpc}$ for the core region of the galaxies in our sample.
We do not find any significant correlation between the $h_4$ radial gradient and the velocity dispersion slope for the sample as a whole.
Even after excluding the outlier NGC~890, the $p$-value of the significance of correlation for a linear fit to the data (with slope $0.34 \pm 0.19$) is $0.17$.
However, we note that 14 out of 19 galaxies with negative $\gamma_{\rm 1 kpc}$ have positive $h_4$ gradients (and positive $h_4$ values in general, as shown in Fig.~\ref{fig:avg_h4_gmos_vp}).

For galaxies without central black holes, the information about $h_4$ can be used to infer the velocity dispersion anisotropy: at small radii close to the galaxy center, various types of models predict that tangential anisotropy produces a peaked LOSVD ($h_4>0$), while radial anisotropy gives a flat-top LOSVD ($h_4<0$) (\citealt{benderetal1994} and references therein).
However, using information about $h_4$ alone to estimate the effect of a black hole on the LOSVD and/or the anisotropy profile is not as straightforward.
For example, \cite{Baesetal2005} studied a two-parameter family of isotropic models with a central black hole.
They found that the $h_4$ profile is significantly affected by the presence of a central black hole for the isotropic models with shallow central density profiles (i.e. the central regions show positive $h_4$ peaks), but didn't see any effects for models with steep density cusps.
Additionally, there are several cases of SMBHs with measured negative $h_4$ where Schwarzschild orbit modeling predicts tangential anisotropy in the galaxy center (\citealt{Pinkneyetal2003, Gebhardtetal2003, mcconnelletal2012}).
This behavior was also reproduced by merger simulations of dynamical systems with central SMBHs that have produced cases with central tangential anisotropy and negative $h_4$, particularly in the systems with higher SMBH masses \citep{Rantalaetal2018}.

Comparing to the isotropic models of \cite{Baesetal2005}, the fact that our galaxies do not exhibit central positive $h_4$ peaks suggests that, if the data resolve the black hole SOI, then the galaxies are unlikely to be isotropic.
The simulations of \cite{Rantalaetal2018} further suggest that central tangential anisotropy could explain the low observed $h_4$, but detailed dynamical modeling is needed to make any robust statements about the behavior of the velocity anisotropy in the center of galaxies.


\section{Jeans modeling}
\label{sec:jam_model}

The stellar kinematics over the large radial coverage of the combined GMOS and Mitchell datasets provide
powerful constraints on the mass components in ETGs.
Here we apply the computationally simple method of Jeans Anisotropic Modeling (JAM; \citealt{cappellari2008}) to
illustrate how the two datasets can be used to infer the central black hole mass, the stellar mass-to-light ratio and the dark matter halo content
of the regular fast rotator NGC~1453 in our sample.

We choose NGC~1453 due to its regular photometric properties and almost purely elliptical isophotes (Fig.~\ref{fig:mge}; \citealt{Goullaudetal2018}). In addition to being a regular fast-rotator, \cite{Eneetal2018} showed that the photometric and kinematic position angles of NGC~1453 are almost perfectly aligned, suggesting that the galaxy may be modeled as an axisymmetric system.
As an interesting note, NGC~1453 also contains a large ($R \sim 20''$) warm ionized gas disc that is rotating almost perpendicularly to the stellar kinematic axis, suggesting that the gas accreted through external processes \citep{Pandyaetal2017}.

The rest of our sample is either slowly rotating and/or shows kinematic and photometric twists. The orbit superposition model \citep{schwarzschild79} is likely a more suitable method for determining the dynamical masses of these galaxies.  Orbit modeling results are beyond the scope of the present paper and will be reported in future work.  

\subsection{Mass Model}

\begin{figure}
  \centering
  \begin{minipage}[b]{0.45\textwidth}
    \includegraphics[width=\textwidth]{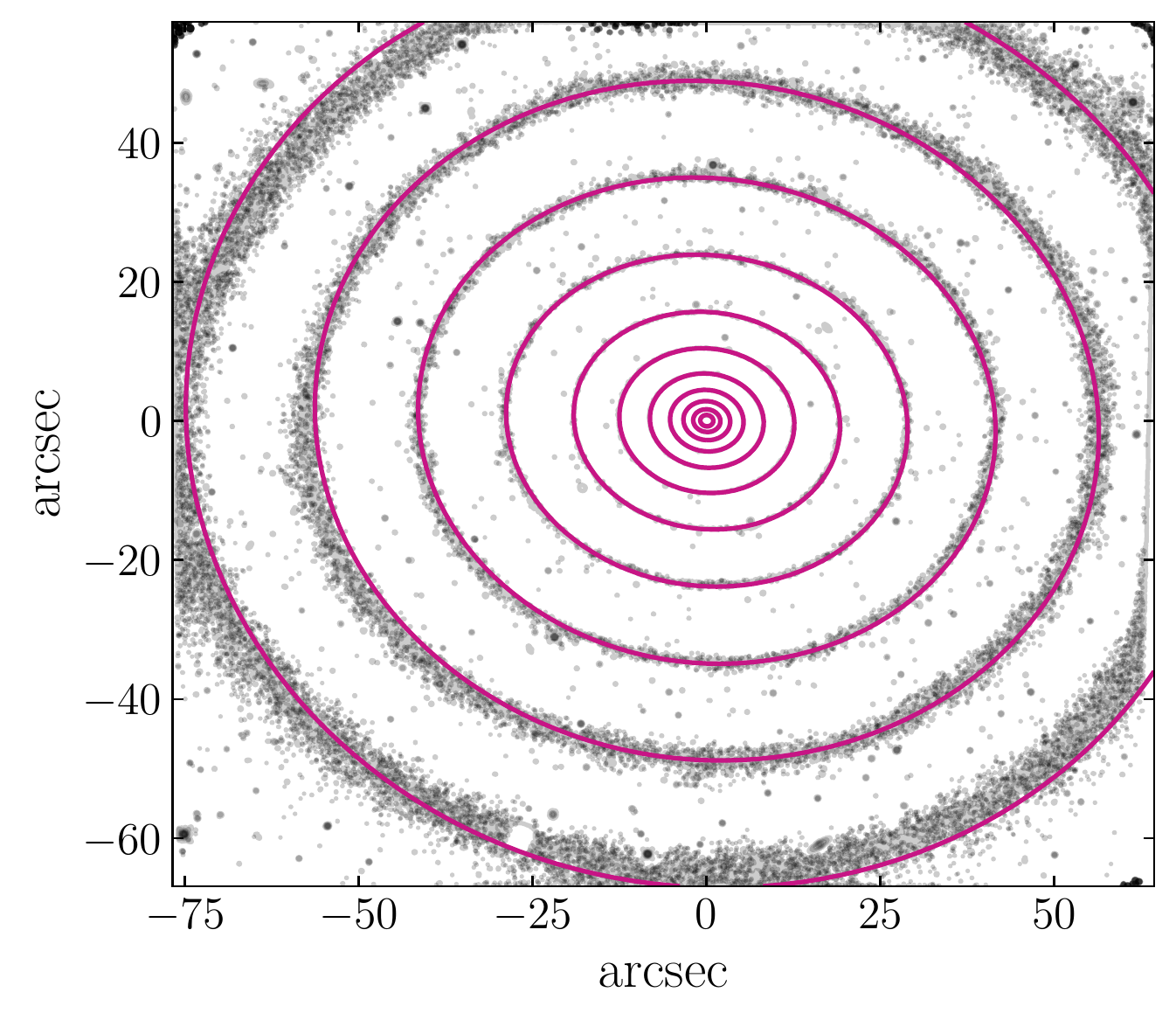}
  \end{minipage}
  \hfill
  \begin{minipage}[b]{0.45\textwidth}
    \includegraphics[width=\textwidth]{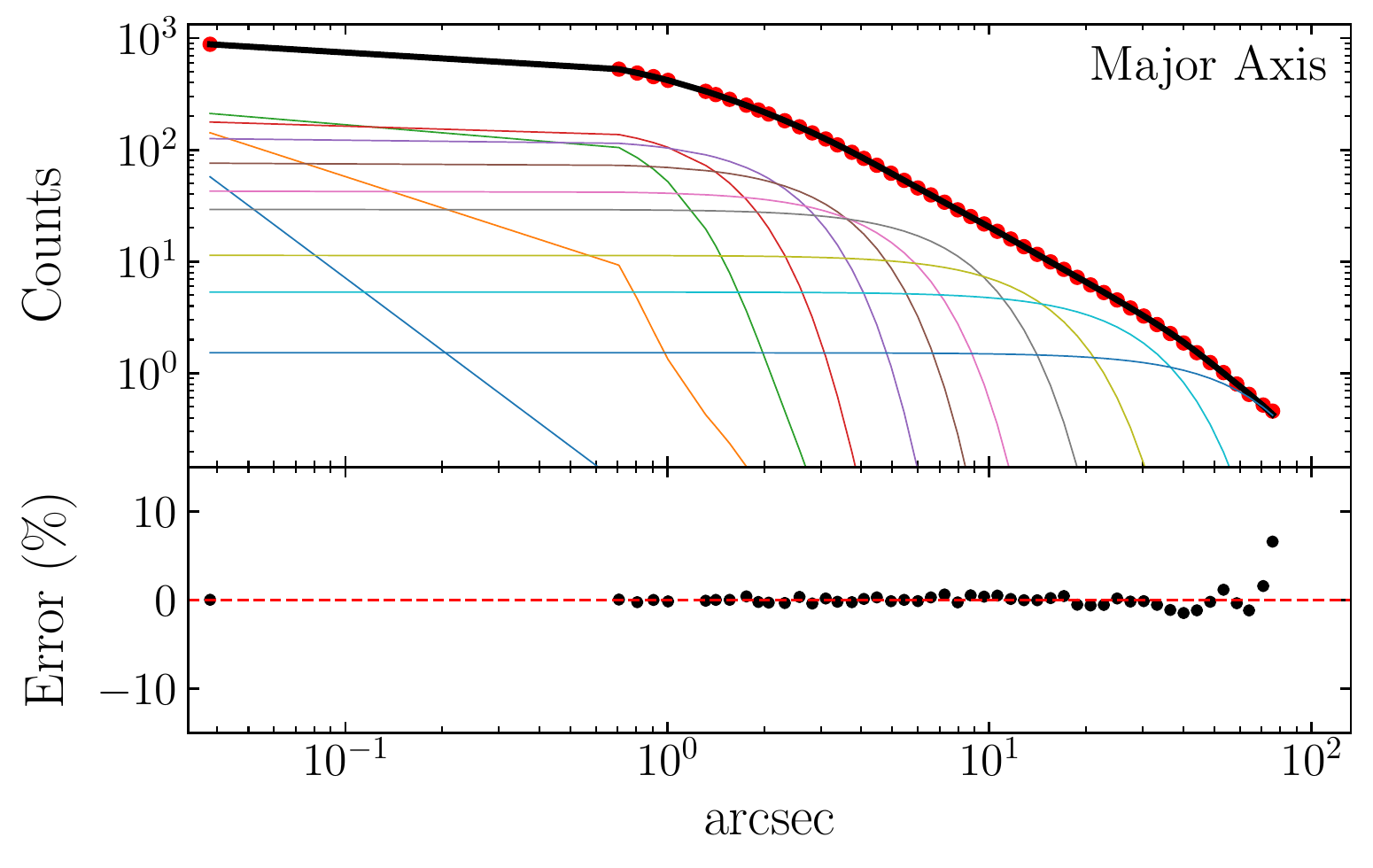}
  \end{minipage}
  \caption{HST WFC3 IR photometry of NGC~1453 (black) and the best-fit MGE model (magenta).  The isophotes have
  no measurable deviation from purely elliptical contours (top panel; \citealt{Goullaudetal2018}). The surface brightness profile is well fit by the sum of 11 Gaussians with small fitting errors (bottom panel).
  \label{fig:mge}}
\end{figure}

We assume that the mass in each galaxy consists of three parts: a central black hole, a stellar mass component and a dark matter halo. To model the stellar component we use the Hubble Space Telescope Wide Field Camera 3 (WFC3) IR photometry in the filter F110W from \cite{Goullaudetal2018}. 
The surface brightness is fitted using the Multi-Gaussian Expansion (MGE) method (\citealt{emsellemetal1994}; \citealt{cappellari2002})
with
a sum of 2-dimensional Gaussian components that share a common center and position angle:
\begin{equation} \label{eg:mge}
    \Sigma (x^\prime, y^\prime) = \sum_{k=1}^{N} \frac{L_k}{2\pi \sigma_k ^2 q_k^\prime } {\rm exp} \Big[-\frac{1}{2 \sigma_k^2}\Big(x^{\prime 2} + \frac{y^{\prime 2}}{q_k^{\prime 2}}\Big)\Big],
\end{equation}
where $x^\prime$ and $y^\prime$ are projected coordinates measured from the galaxy center, with $x^\prime$ being along the photometric major axis. The subscript $k$ labels the individual Gaussian components, $L_k$, $\sigma_k$, $q_k^{\prime}$ are the luminosity, dispersion and projected axis ratio of each Gaussian respectively. To obtain the MGE fit, the model predictions need to be convolved with the point spread function (PSF), which is taken from \cite{Goullaudetal2018}, and is also expressed as an MGE using 5 nearly-circular Gaussian components (with axis ratios $>0.98$).
Our best-fit MGE to the surface brightness of NGC~1453 consists of 11 Gaussian components, which are summarized in Table \ref{tab:mge} and plotted in Fig.~\ref{fig:mge}.
The small fitting residuals (bottom panel) demonstrate that the MGE model agrees very well with the data. 

\begin{table}
\centering
\caption{ Gaussian components of the MGE fit to the HST WFC3 IR photometry of NGC~1453 (shown in Figure \ref{fig:mge}). All 11 Gaussians share the same center and position angle of 28.5 degrees. From left to right: central surface density $I_k = L_k/2\pi \sigma_k ^2 q_k^\prime$ (calculated using an absolute solar magnitude $M_{\odot{\rm, F110W}}=4.54$
), dispersion $\sigma_k$ in arcseconds and axis ratio $q_k^{\prime}$ of each Gaussian component.   \label{tab:mge} } 
\begin{tabular}{c|c|c}
$I_k \ \ [L_{\odot} / {\rm pc}^{2}]$ & $ \ \ \sigma_k \ \ [''] \ \ $ & $ \ \ q_k^{\prime} \ \ $ \\
\hline
& & \\
 $23280.1$  &  $ \ 0.041 \ $   &     $ \ 0.891 \ $ \\
 $9851.1$   &    $0.270$    &     $0.882$ \\ 
 $12303.7$  &    $0.575$    &     $0.900$ \\
 $9798.8$  &    $0.963$    &     $0.816$ \\  
 $6734.01$  &    $1.595$    &     $0.800$ \\
 $4000.64$  &    $2.370$    &     $0.865$ \\
 $2235.06$  &    $3.403$    &     $0.800$ \\  
 $1526.44$  &    $5.743$  &       $0.835$ \\
 $591.08$   &    $10.292$   &     $0.800$ \\
 $277.19$    &    $20.778$   &     $0.821$ \\
 $79.34$  &    $47.289$   &     $0.897$ \\

\end{tabular}
\end{table}

\begin{figure*}
\centering
 \includegraphics[scale = 0.47]{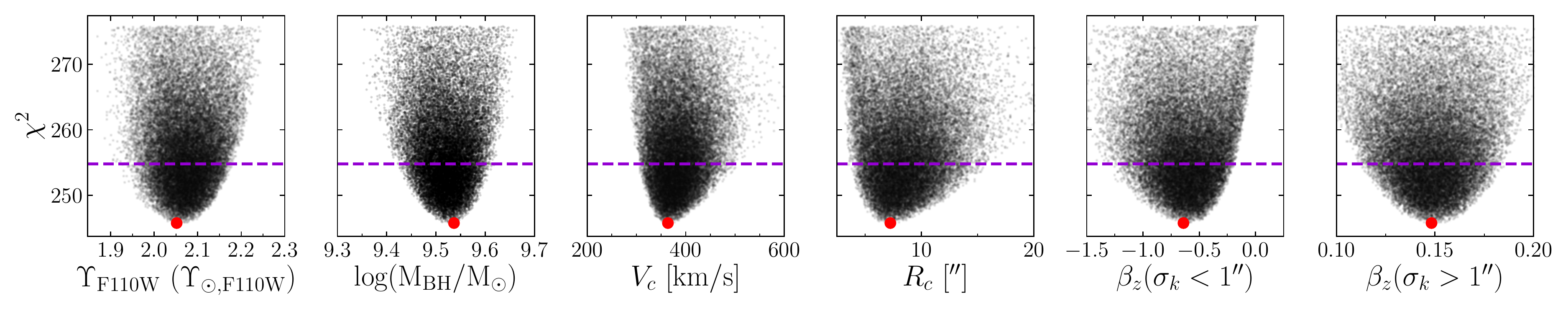}
   \caption{Results of JAM model fit to the GMOS and Mitchell stellar kinematics for NGC~1453.  The $\chi^2$ is shown for each of the six mass model parameters. Each black point represents a single mass model. The best-fit model (red circle) has $\Upsilon_{\rm F110W} = 2.06 \ \Upsilon_{\rm \odot, F110W }$, $M_{\rm BH} = 3.3 \times 10^9 M_{\odot}$, $V_c = 364 \ {\rm km/s}$, $R_c = 7.2 \ {\rm kpc}$, $\beta_z^G (\sigma_k < 1'') = -0.58$ and $\beta_z^G (\sigma_k > 1'') = 0.15$.  The dashed purple line indicates $\Delta \chi^2 = 9$, which is the $3 \sigma$ confidence level for one parameter. \label{fig:chi2} }
 \end{figure*}

Given an inclination and stellar mass-to-light ratio $\Upsilon_{\rm F110W}$, the 2-dimensional MGE fit can be deprojected and converted into a realistic 3-dimensional stellar mass density profile. Throughout this work we assume an edge-on orientation (inclination of $90^{\circ}$) and a spatially constant $\Upsilon_{\rm F110W}$.

We parametrize the dark matter halo using a two-parameter logarithmic potential:
\begin{equation}\label{eq:dm_pot}
    \Phi_{\rm DM}(r) = \frac{1}{2} V_c ^2 {\rm ln}\Big(r^2 + R_c^2\Big), 
\end{equation}
where $R_c$ is the characteristic radius of the halo potential and $V_c$ is the circular speed at $r \rightarrow \infty$ (such a DM description was, e.g., favored by orbit-based models of M87 in \citealt{murphy2011}). The logarithmic potential corresponds to a density profile given by (\citealt{binneytremaine2008}):
\begin{equation}\label{eq:dm_pot}
    \rho_{\rm DM}(r) = \frac{V_c^2}{4\pi G} \frac{3 R_c^2 + r^2}{(R_c^2 + r^2)^2}. 
\end{equation}
Before being passed to JAM, the dark matter density profile is fitted using a 1-dimensional MGE model. 

\subsection{Modeling velocity anisotropy}

We augment the four parameters describing the mass profile with a description for the anisotropy,
\begin{equation}
\beta_z = 1 -\frac{\overline{v_z^2}}{\overline{v_R^2}},
\end{equation}
which quantifies the flattening of the velocity ellipsoid along the minor axis. 

Previous works with JAM have typically assumed a globally constant $\beta_z$, although there have been some attempts at introducing moderate spatial variation, by having $\beta_z$ vary between the different Gaussian components of the MGE (e.g., \citealt{cappellari2015}). The large radial span of our combined two data sets motivates including at least some spatial dependence. Perhaps most importantly, our model should distinguish between the core and the outskirts of the galaxy. The core-like part of the light profile of massive galaxies is commonly interpreted as being the result of black hole scouring (\citealt{faber97}), which in turn predicts a bias towards tangential orbits in the centre and more radial orbits in the outskirts (\citealt{quinlan97}; \citealt{MilosavljevicMerritt2001}). In order to at least partially replicate orbit-type variation, we assign separate anisotropies $\beta_z^G$ to the Gaussians with $\sigma_k < 1''$ and the Gaussians with $\sigma_k > 1''$. The choice of $1''$ is motivated by the fact that the light profile of NGC~1453 starts to fall off more steeply once $R \gtrsim 1''$ (see bottom panel of Fig.~\ref{fig:mge}). 
 
\subsection{Fitting the kinematic data}

Our mass model is specified by six parameters: $M_{\rm BH}$, $\Upsilon_{\rm F110W}$, $R_c$, $V_c$, $\beta_z^G (\sigma_k < 1'')$ and $\beta_z^G (\sigma_k > 1'')$. 
JAM places constraints on these parameters by comparing the model predictions for $v_{\rm rms} = \sqrt{V^2 + \sigma^2}$ (including instrumental PSF convolution) to the observed $v_{\rm rms}$ for each Voronoi bin. 
We use the GMOS PSF (modeled as a Gaussian with dispersion $0.297''$) since the PSF is most important near the center, at the locations of the GMOS data points.
We compute the PSF by fitting a Gaussian function to point sources from the GMOS acquisition images, which are usually taken before a group of 3 -- 4 science exposures.
The final PSF estimate is a weighted average of the Gaussian FWHMs, weighed by the total exposure time of the group of science exposures that each acquisition image precedes.
In the fit, we exclude the few Mitchell kinematic points with $R < 5''$ since this region is better measured by GMOS. 

We find the JAM parameters by first running a broad regular-grid search, followed by a best-fit estimation using Bayesian inference. We calculate the posterior probability distribution using an implementation (\citealt{cappellari2013}) of the Adaptive Metropolis algorithm of \cite{haario2001}. The best-fitting model (with 3$\sigma$ errors) is $\Upsilon_{\rm F110W} = 2.06 \pm 0.13 \ \Upsilon_{\rm \odot, F110W }$, $M_{\rm BH} = 3.29 \pm 0.75 \times 10^{9} \ M_{\odot}$, $V_c = 364 \pm 134 \ {\rm km/s}$,  $R_c = 7.2 \pm 8.3 \  {\rm kpc}$, $\beta_z^G (\sigma_k < 1 '' ) = -0.58 \pm 0.62$ and $\beta_{z}^G (\sigma_k > 1'') = 0.15 \pm 0.04$.

\begin{figure}
\centering
 \includegraphics[scale = 0.52]{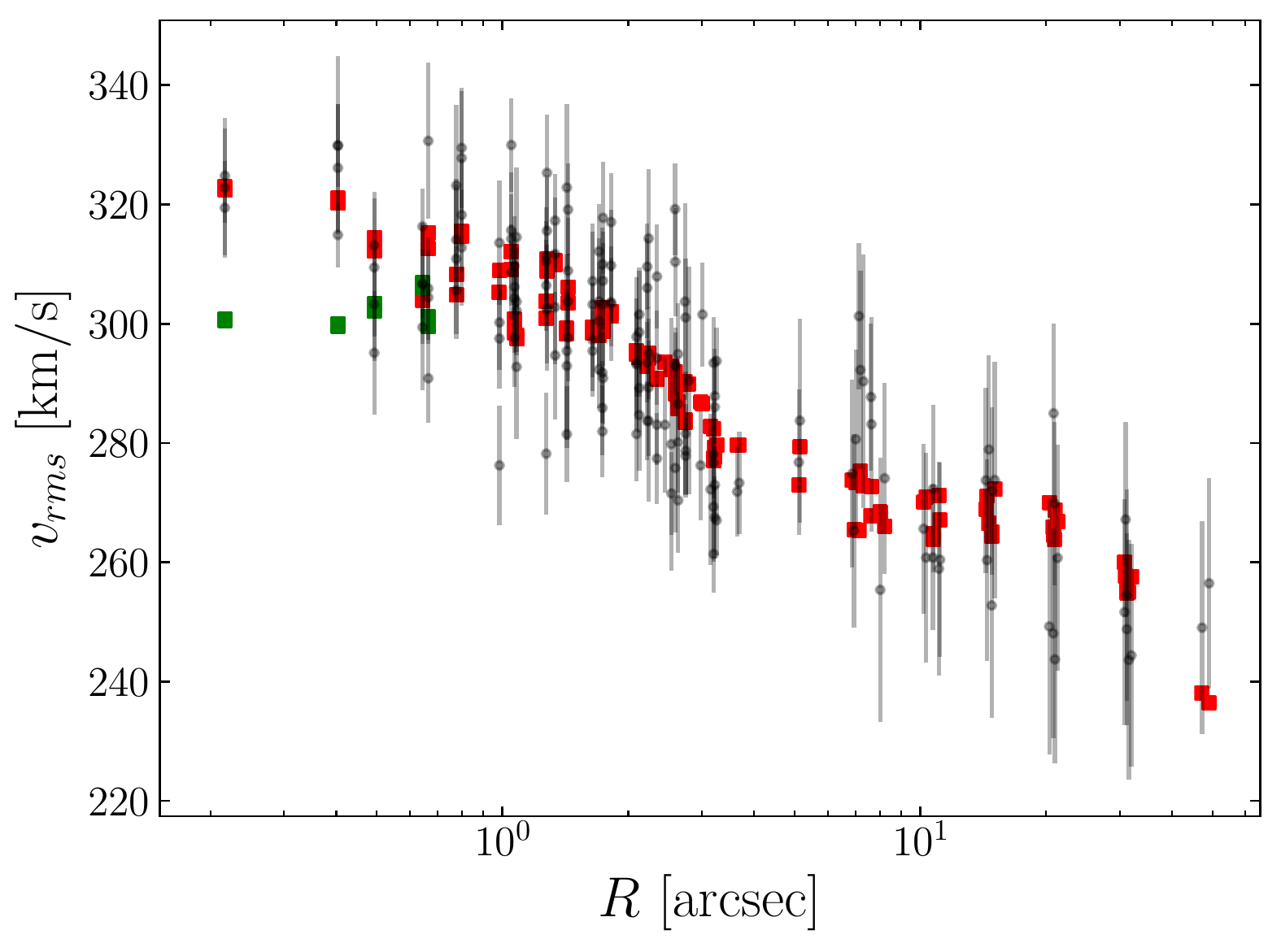}
   \caption{Line-of-sight rms velocity, $v_{\rm rms}=\sqrt{V^2 + \sigma^2}$, of NGC~1453 measured from GMOS and Mitchell IFS (black points with error bars) versus the best-fit JAM predictions (red squares).
   The GMOS data points within $\sim 1''$ 
      are well fitted by a $3.3 \times 10^9 M_{\odot}$ black hole, whereas a model without a central black hole (green squares) fails to match the kinematics at small radii.
     \label{fig:fit} }
 \end{figure}

The $\chi^2$ of a large suite of JAM models 
shows a clear minimum for each of the six mass model parameters
(Fig.~\ref{fig:chi2}).
The constraints on the two dark matter halo parameters
are weaker primarily because at $r \ll R_c$, $\rho_{\rm DM} \propto V_c^2 / R_c^2$. As a result, there is significant degeneracy between the two halo parameters. The negative $\beta_z^G (\sigma_k < 1 '')$ and positive $\beta_z^G (\sigma_k > 1 '')$  of our best-fit model indicate a radially increasing $\beta_z$ profile (the $\beta_z$ profile that can be obtained from the best-fit $\beta_z^G$ is not a step function, but smoothly increasing from $\beta_z \approx -0.25$ at small radii to $\beta_z \approx 0.15$ outside of the core) \footnote{We also ran JAM with the two-component $\beta_z$ replaced by a single, constant $\beta_z$. The resulting best-fit model ($\Upsilon_{\rm F110W} = 2.25 \ \Upsilon_{\rm \odot, F110W } $, $M_{\rm BH} = 2.27 \times 10^{9} \ M_{\odot}$, $V_c = 404 \ {\rm km/s}$, $R_c = 11.6 \  {\rm kpc}$, $\beta_z = 0.11$) produces a worse fit, with $\chi^2 = 292.7$.}. JAM's $\beta_z$ cannot be directly related to tangential/radial anisotropy (defined in spherical coordinates), but our data clearly favors a velocity anisotropy that is different in the inner and outer parts of the galaxy.  
Figure \ref{fig:fit} shows our best-fitting $v_{rms}$ predictions, demonstrating that the trends in both the GMOS ($R < 5 ''$) and Mitchell ($R > 5 ''$) data sets can be well reproduced by JAM
and a model without a central black hole fails to reproduce the observed kinematics at small radii.

While JAM is plausibly a suitable method for NGC~1453 and other fast rotators, the utility of JAM is limited by its assumptions of axisymmetry, cylindrically aligned velocity ellipsoids and by the issue that its solutions could be unphysical. These assumptions are not well motivated in slowly rotating, triaxial ETGs, which constitute the majority of our galaxies.  Furthermore, the MGE-based anisotropy profiles, which are globally constant or with limited spatial variation, may be too simplistic for representing a galaxy's true anisotropy structure.
Future papers in the MASSIVE survey series will present results from full-scale dynamical mass modeling using stellar orbit libraries that take advantage of the full set of kinematic moments from GMOS and Mitchell observations (Liepold et al. in preparation; Quenneville et al. in preparation).

\section{Conclusions}
\label{sec:conclusions}

We have presented the first results from the high spatial-resolution spectroscopic component of the MASSIVE survey.
The spatially-resolved spectroscopic observations were obtained with the Gemini GMOS-N IFS for the central $5'' \times 7''$ (about 2 kpc across) regions of 20 ETGs in the MASSIVE survey.  
These galaxies have $M_K \leq -25.5$ ($M_* \gtrsim 10^{11.7} M_\odot $) and are located at a median distance of $\sim 70$ Mpc.  
We measured the stellar kinematics from high-$S/N$ ($\sim 120$) spectra using a Gauss-Hermite parameterization of the LOSVD.
We obtained two-dimensional maps of the first four kinematic moments ($V$, $\sigma$, $h_3$, and $h_4$) and compared them to the large-scale stellar kinematics obtained from the Mitchell IFS (\citealt{Vealeetal2017a, Vealeetal2017b, Vealeetal2018}).  The two IFS datasets together cover a length scale of $\sim 0.3''-100''$, or $\sim 0.1-30$ kpc.  Our main findings are as follows:

\begin{itemize}
    \item The velocity maps of most galaxies in the sample show some level of rotation in the central regions (upper left panel in Figs.~\ref{fig:A1}-\ref{fig:A10} of Appendix~\ref{sec:kinematic_moments_plots}).
    Three galaxies (NGC~1129, NGC~1453, and NGC~2693) exhibit fast rotations with $|V|$ rising up to $\sim 100 - 150$ \kms\ within 1 kpc, and a fourth galaxy (NGC~1700) has a counter-rotationg core that rotates in the opposite direction of the main body rotation (Fig.~\ref{fig:v-profiles} and Fig.~\ref{fig:1700-vmap}).
    These 4 galaxies also have noticeable rotation out to one effective radius, $\lambda_e \gtrsim 0.1$.
    The rest of the galaxies have central rotations (if detected) with $|V| \lesssim 30$ \kms, as well as low velocity rotation up to $R_e$ ($\lambda_{\rm 1 kpc}, \lambda_e \lesssim 0.1$).
    The kinematic axis of the central rotation is not necessarily aligned with the photometric axis or the large-scale kinematic axis, indicating a diverse merger history within this sample of 20 high-mass ETGs.
    
    \item 
    The velocity dispersion $\sigma$ within $1$ kpc reaches $\sim 300$ \kms\ or beyond in 6 of the 20 galaxies and is greater than 250 \kms\ for 14 galaxies.
    We measure the luminosity-weighted average $\sigma$ within an aperture of radius $R_e/8$ versus $R_e$ and find an aperture correction relation of $\sigma_{e/8}/\sigma_e = 1.044 \pm 0.078$ (Fig.~\ref{fig:aperture_correction}).

    \item 
    For all but one galaxy, the velocity dispersion profiles $\sigma(R)$ in the radial range of $\sim 0.3''-3''$ are well fit by a single power law form with \textit{negative} logarithmic slope (Fig.~\ref{fig:sigma_profiles}).
    A rising $\sigma(R)$ towards smaller radii is indicative of the presence of a central SMBH but can also be caused by central radial anisotropy in the stellar velocities.
    On large scales, by contrast, the logarithmic slope of $\sigma(R)$ at $1 R_e$ ranges from $-0.25$ to $+0.25$, reflecting varying degrees of contributions from dark matter mass as well as velocity anisotropies.  We will use the observed higher-order velocity moment $h_4$ to break this mass-anisotropy degeneracy in future dynamical mass modeling work.
    
    \item For the galaxies with clear rotation in the inner $\sim 1$ kpc, the spatially-resolved skewness $h_3$ in this region is anti-correlated with the velocity, i.e., the spatial bins with higher $V$ values have more negative $h_3$.
    The slope of this correlation, $\Delta h_3/\Delta (V/\sigma)$, ranges between $-0.1$ and $-0.25$ for the galaxies with relatively high spins in our sample ($\lambda_{\rm 1 kpc} \gtrsim 0.1$).
    The $h_3$-$V$ anti-correlation within a galaxy indicates disc-like kinematics in both the core and  main body of fast-rotating massive ETGs.
    
    \item 
    The kurtosis $h_4$ is generally positive, with
    14 of the 20 galaxies having a positive average $h_4$ within $1$ kpc, and 18 of them having positive average $h_4$ within $1 R_e$ (Fig.~\ref{fig:avg_h4_gmos_vp}).
     Most galaxies also show a rising or flat radial profile $h_4(R)$.
     In the absence of a central mass,
    peaked LOSVDs ($h_4>0$) at small radii are often associated with tangential anisotropy, which tends to lower the central line-of-sight $\sigma$.  
     Yet 19 out of the 20 galaxies in this study are observed to have $\sigma(R)$ that increase towards the center. The presence of central black holes 
     is a possible explanation.
    We will perform full-scale stellar orbit modeling to determine the velocity structures, stellar and dark matter mass distributions, and black hole masses in these galaxies.
    
    \item 
     Using the GMOS and Mitchell stellar kinematics,
     we applied the Jeans modeling method to
     measure the mass distributions in NGC~1453, the most regular fast rotator in our sample.
      To partially account for spatial variations in the stellar velocity anisotropy, we allow the anisotropy parameter $\beta_{z}$ in the inner and outer parts of the galaxy to be different. The JAM results show that our kinematics point towards a nonzero central black hole mass and a spatially varying velocity anisotropy.
     The best-fit mass model (with 3$\sigma$ errors) has a black hole mass of $3.29 \pm 0.75 \times 10^{9} M_{\odot}$, a stellar mass-to-light ratio of $\Upsilon = 2.06 \pm 0.13 \ \Upsilon_{\odot}$ (in WFC3 F110W band), and a circular velocity of $V_c = 364 \pm 134$ \kms\ for the dark matter halo.
\end{itemize}

High spatial-resolution kinematics that resolve the sphere of influence of the SMBH are a key requirement for any attempt at black hole mass determination through dynamical modeling.
Combining the small scale results presented here with the large-scale kinematics of previous MASSIVE papers will allow us to model the mass distributions of nearby massive ETGs and study their assembly histories.
New SMBH mass measurements will help refine the various correlation relations between black holes and their host galaxy properties.
In particular, the high-mass range sampled by our galaxy sample may be relevant for the exploration of the $M_\bullet - \sigma_e$ saturation at high $\sigma_e$ (e.g., \citealt{laueretal2007, mcconnelletal2011a, mcconnelletal2012, mcconnellma2013, kormendyho2013}).

\acknowledgments
This work is based on observations obtained at the Gemini Observatory processed using the Gemini IRAF package, which is operated by the Association of Universities for Research in Astronomy, Inc., under a cooperative agreement with the NSF on behalf of the Gemini partnership: the National Science Foundation (United States), National Research Council (Canada), CONICYT (Chile), Ministerio de Ciencia, Tecnolog\'{i}a e Innovaci\'{o}n Productiva (Argentina), Minist\'{e}rio da Ci\^{e}ncia, Tecnologia e Inova\c{c}\~{a}o (Brazil), and Korea Astronomy and Space Science Institute (Republic of Korea).
We thank Richard McDermid for his help in the early stages of acquiring and preparing the GMOS data. 
The MASSIVE survey is supported in part by NSF AST-1411945, AST-1411642, AST-1815417 and AST-1817100, HST GO-14210, GO-15265 and AR-14573, and the Heising-Simons Foundation.
J.L.W. is supported in part by NSF grant AST-1814799.
Gemini North is located on Maunakea, a sacred place for indigenous Hawaiians who have honored it before and since the construction of modern astronomical facilities. We hold great privilege and responsibility in using the Maunakea summit.

%

\vspace{5mm}
\facility{Gemini:Gillett (GMOS)}





\appendix

%


\section{Kinematic moments plots}
\label{sec:kinematic_moments_plots}

\begin{figure*}
  \includegraphics[width=\textwidth]{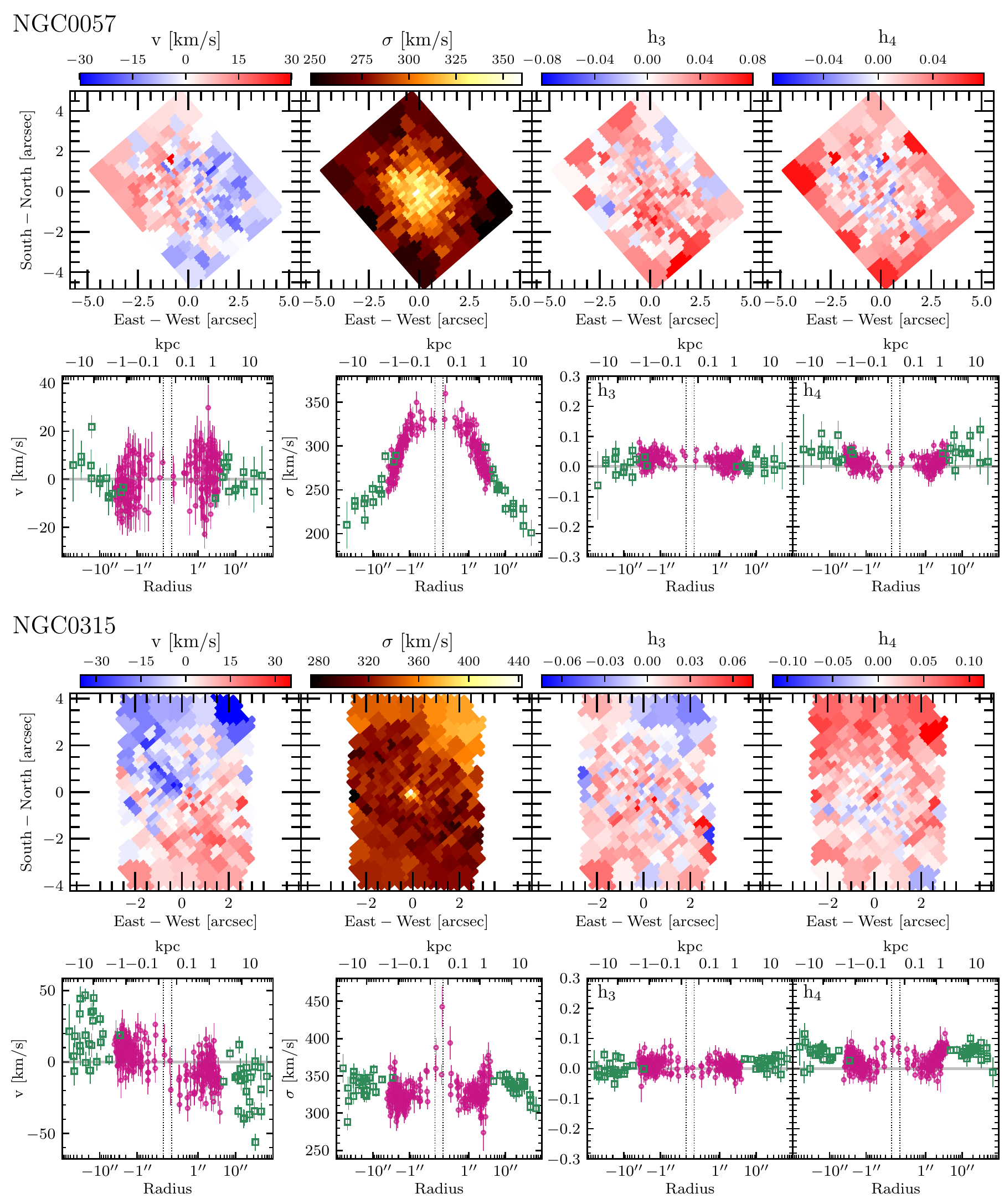} 
\caption{Kinematic moments for NGC~57 and NGC~315. The top row panels show two-dimensional maps of the first four GH moments ($V, \sigma, h_3, h_4$) as measured from the GMOS data. The bottom row panels show two-sided radial profiles from GMOS (purple circles) and Mitchell (green squares) data. The data are unfolded along the IFU PA; points with positive radius are within $\pm 90 \degr$ of the IFU PA. The data are plotted using a logarithmic scale for radius. To guide the eye, vertical dotted lines denote $\pm 0.2''$.}
\label{fig:A1}
\end{figure*}

\begin{figure*}
  \includegraphics[width=\textwidth]{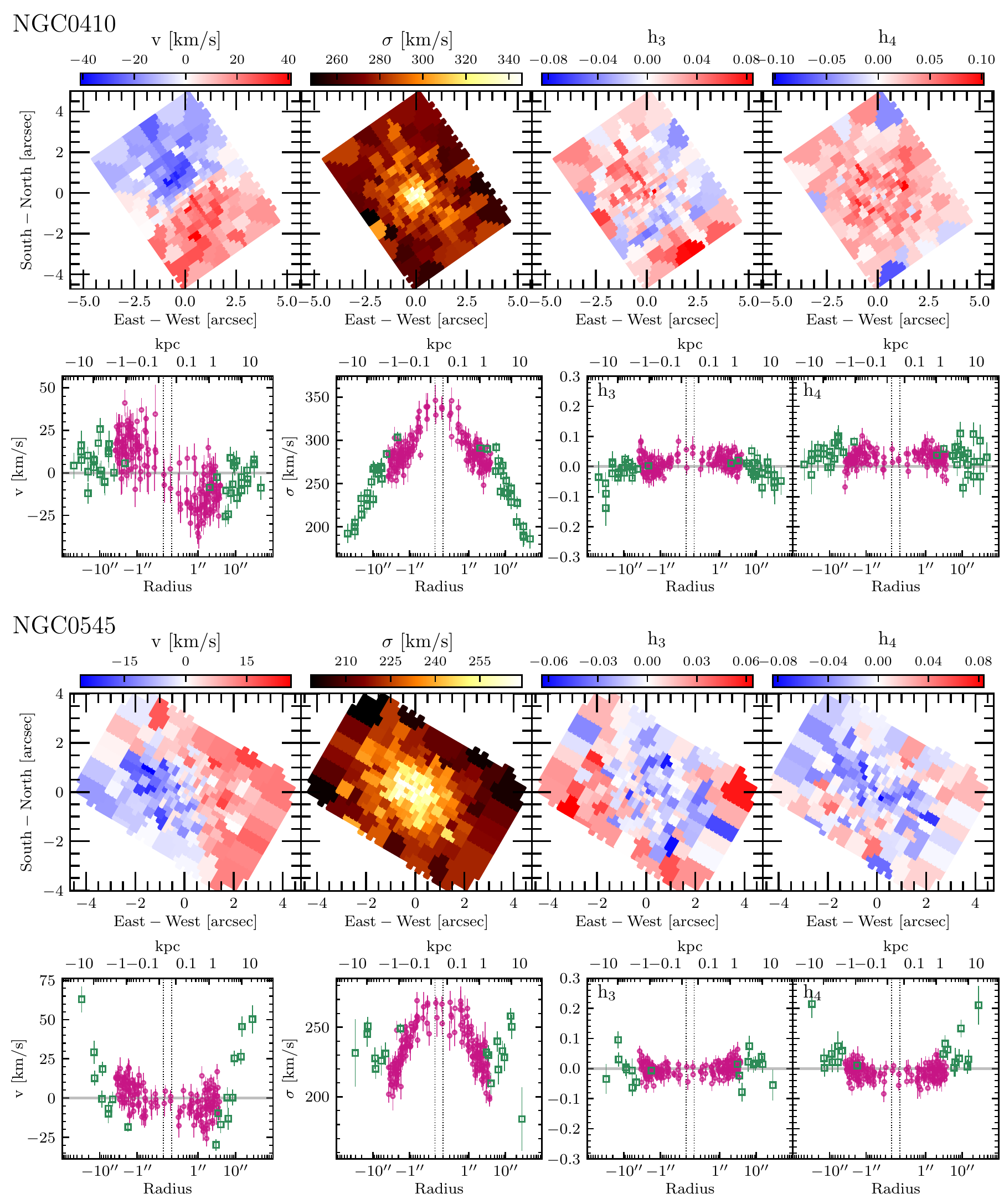} 
\caption{Same as in Fig.~\ref{fig:A1} but for NGC~410 and NGC~545.}
\end{figure*}

\begin{figure*}
  \includegraphics[width=\textwidth]{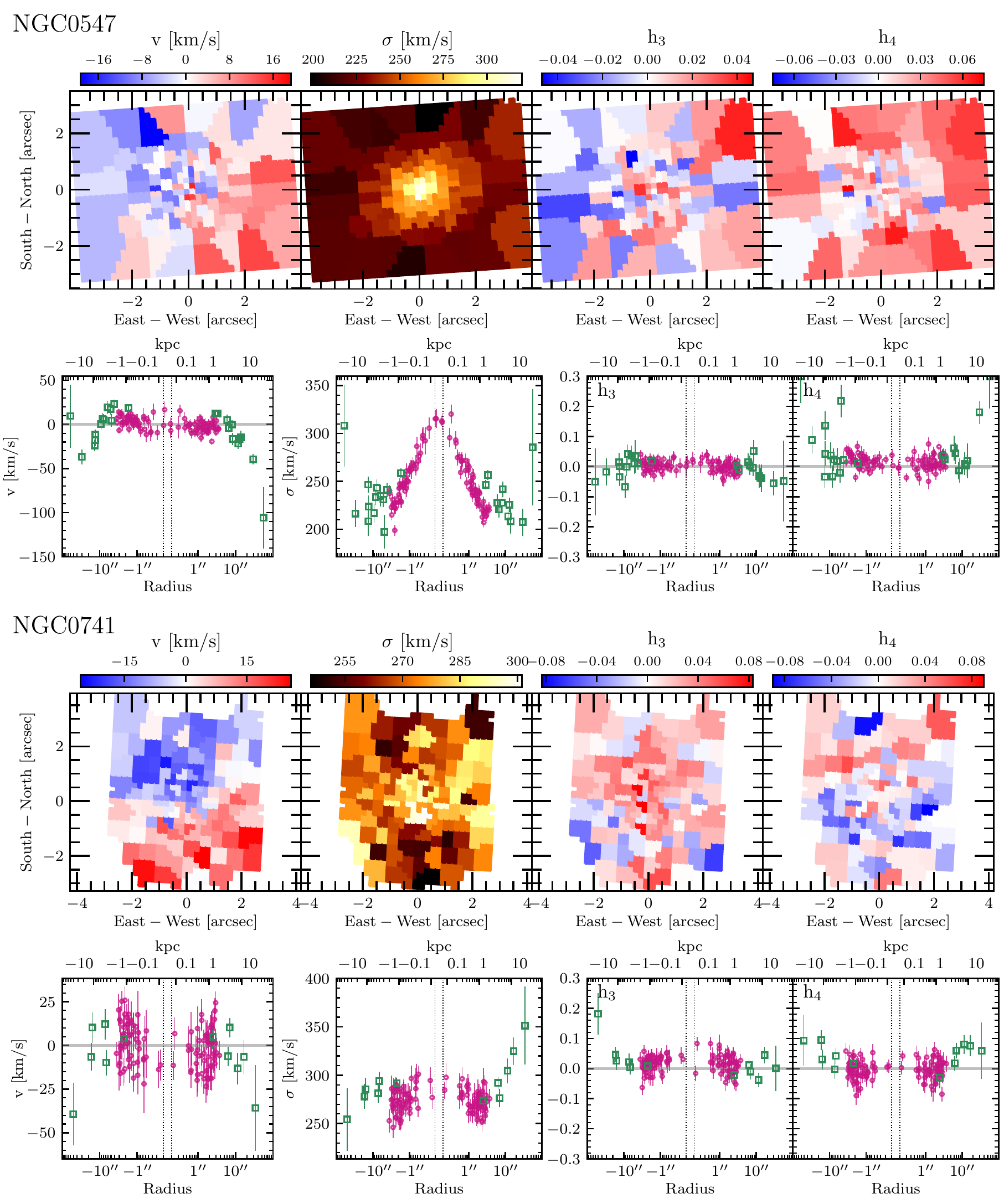}
\caption{Same as in Fig.~\ref{fig:A1} but for NGC~547 and NGC~741.}
\end{figure*}

\begin{figure*}
  \includegraphics[width=\textwidth]{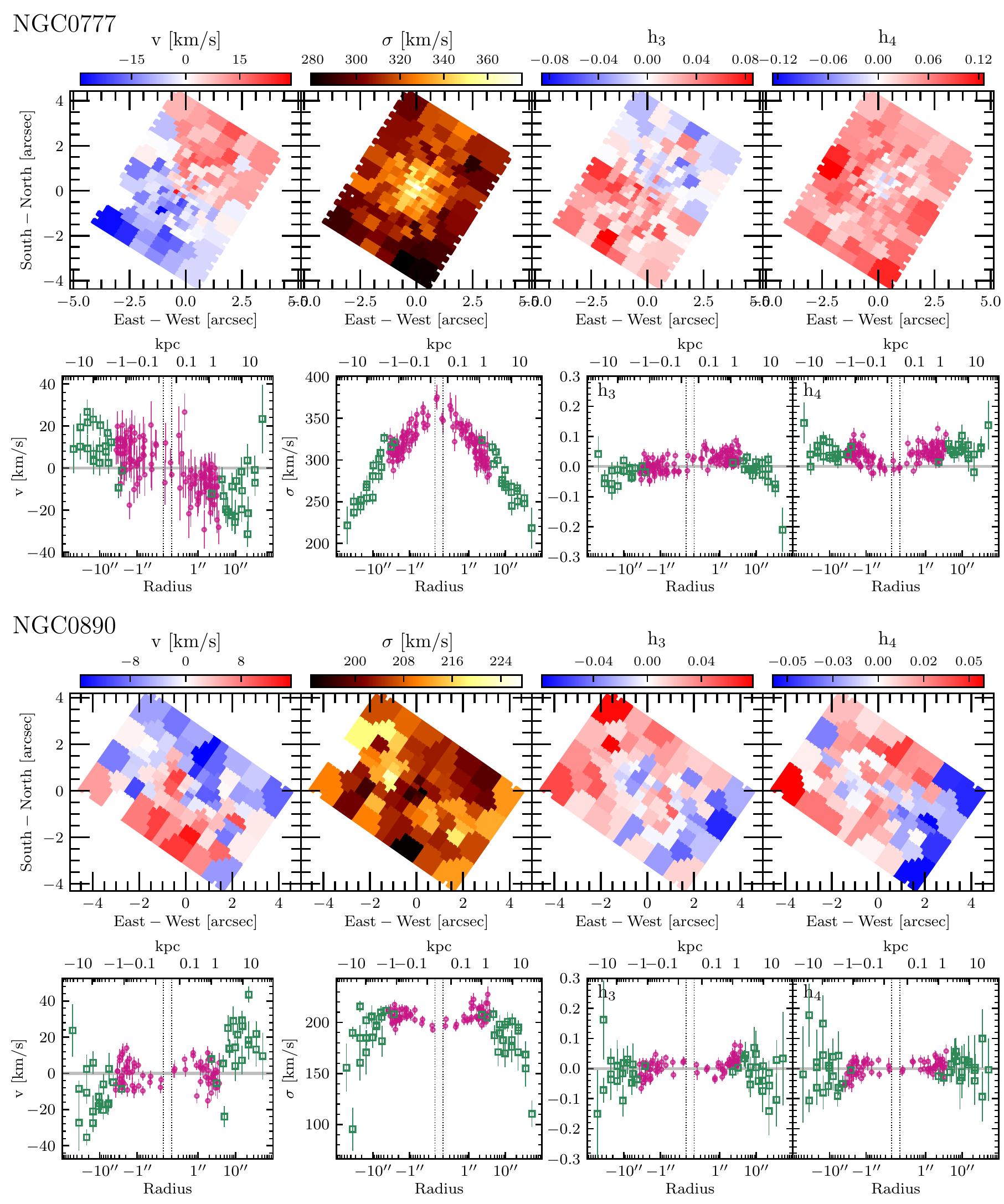} 
\caption{Same as in Fig.~\ref{fig:A1} but for NGC~777 and NGC~890.}
\end{figure*}

\begin{figure*}
  \includegraphics[width=\textwidth]{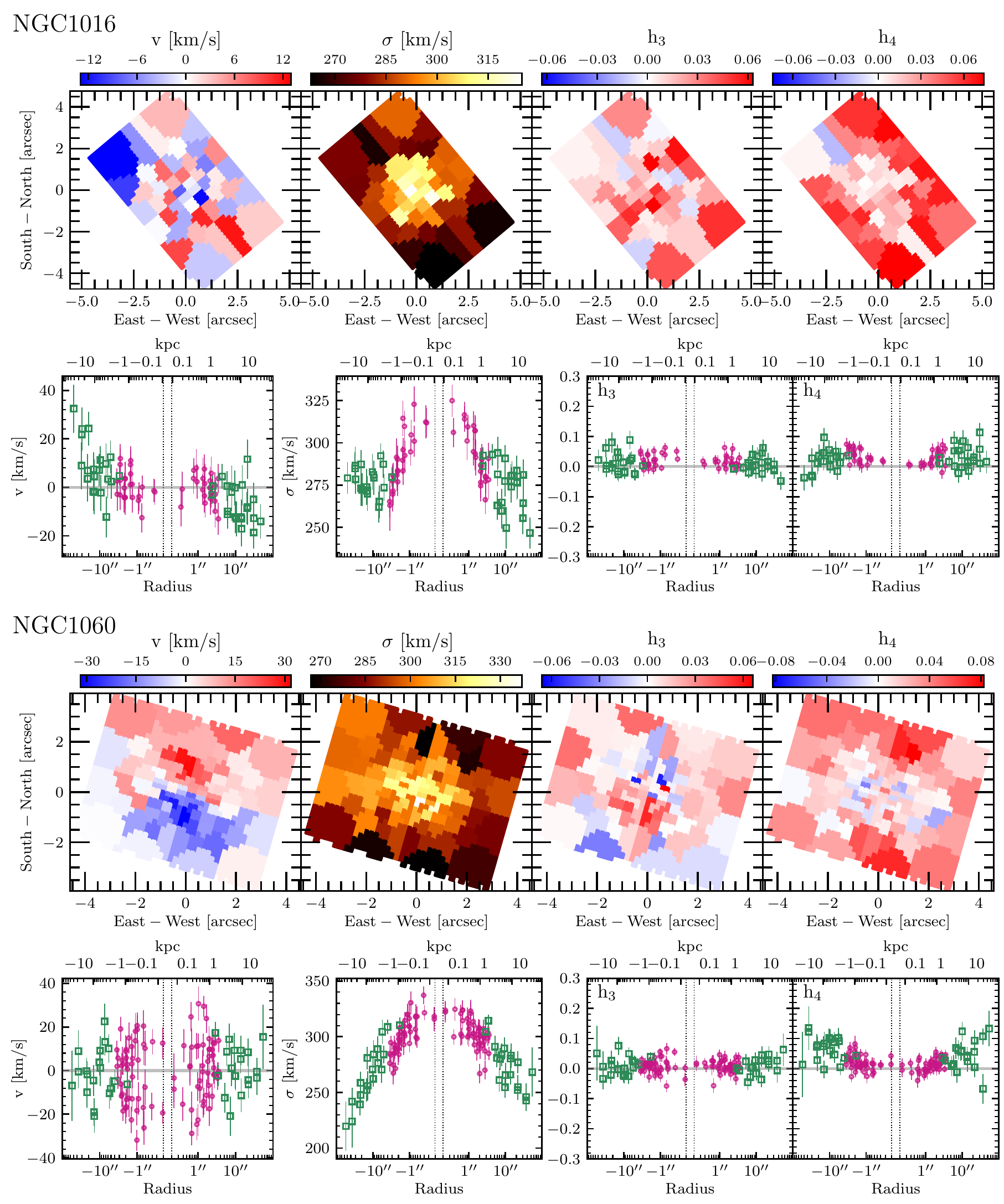} 
\caption{Same as in Fig.~\ref{fig:A1} but for NGC~1016 and NGC~1060.}
\end{figure*}

\begin{figure*}
  \includegraphics[width=\textwidth]{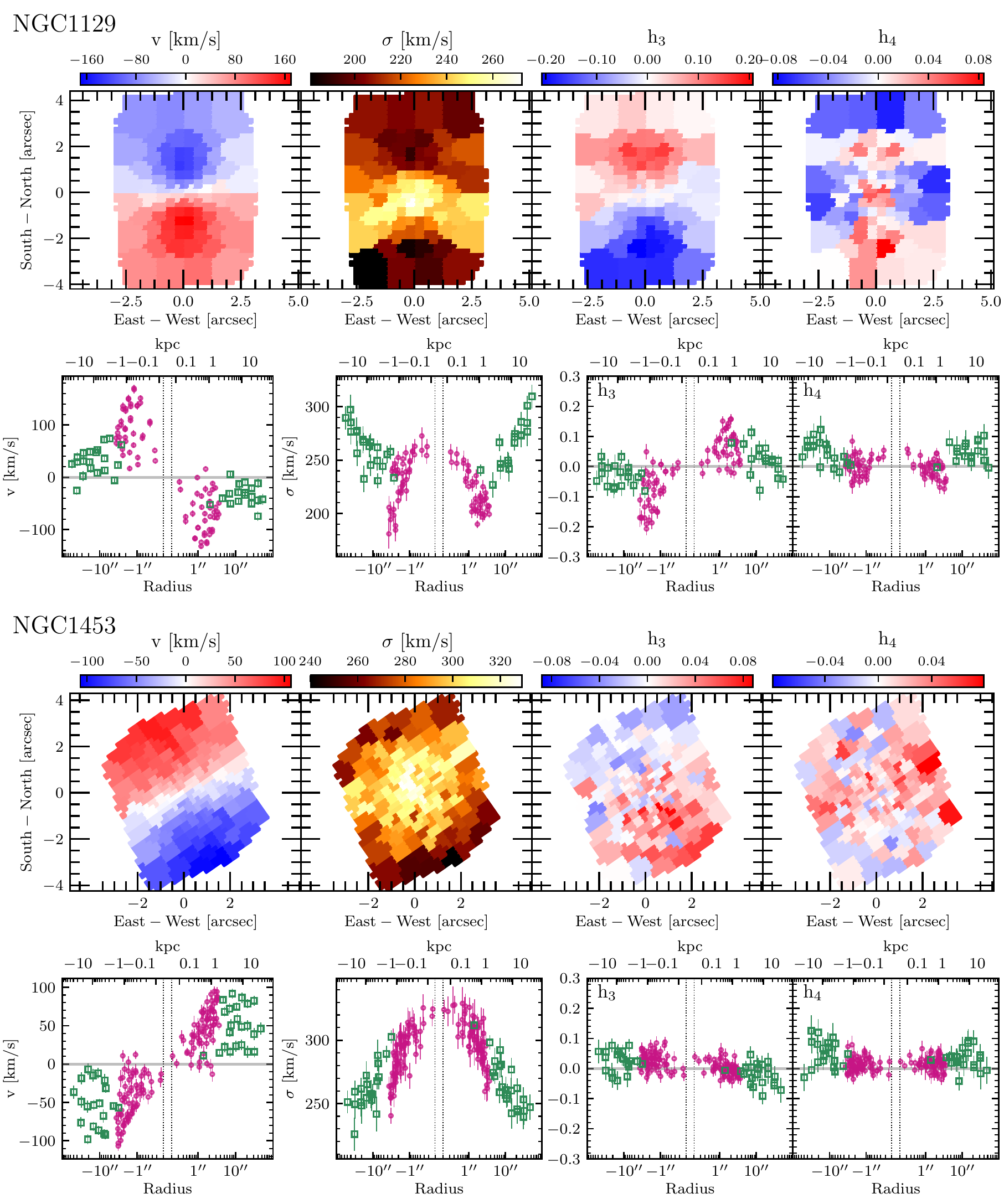} 
\caption{Same as in Fig.~\ref{fig:A1} but for NGC~1129 and NGC~1453.}
\end{figure*}

\begin{figure*}
  \includegraphics[width=\textwidth]{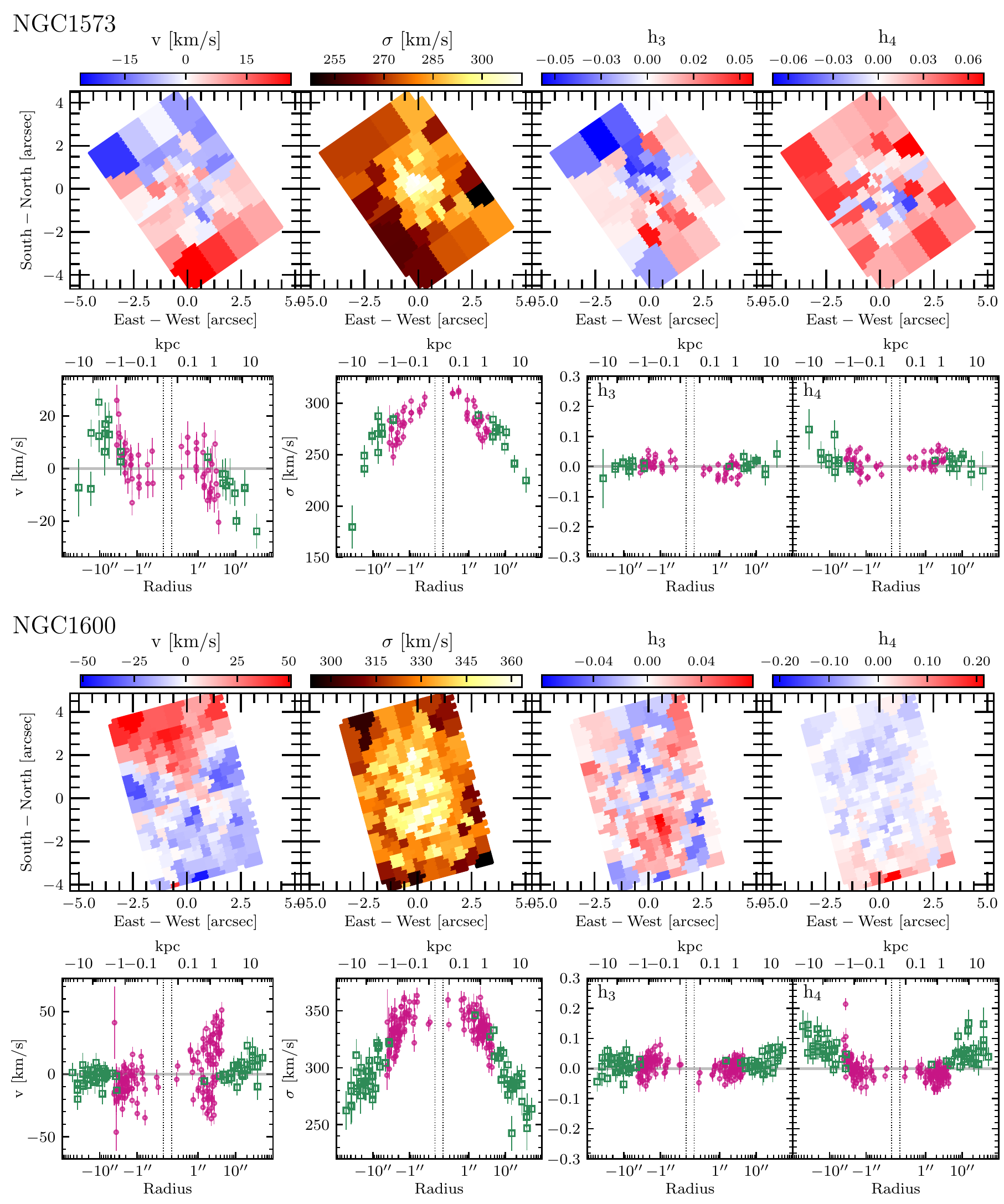} 
\caption{Same as in Fig.~\ref{fig:A1} but for NGC~1573 and NGC~1600.}
\end{figure*}

\begin{figure*}
  \includegraphics[width=\textwidth]{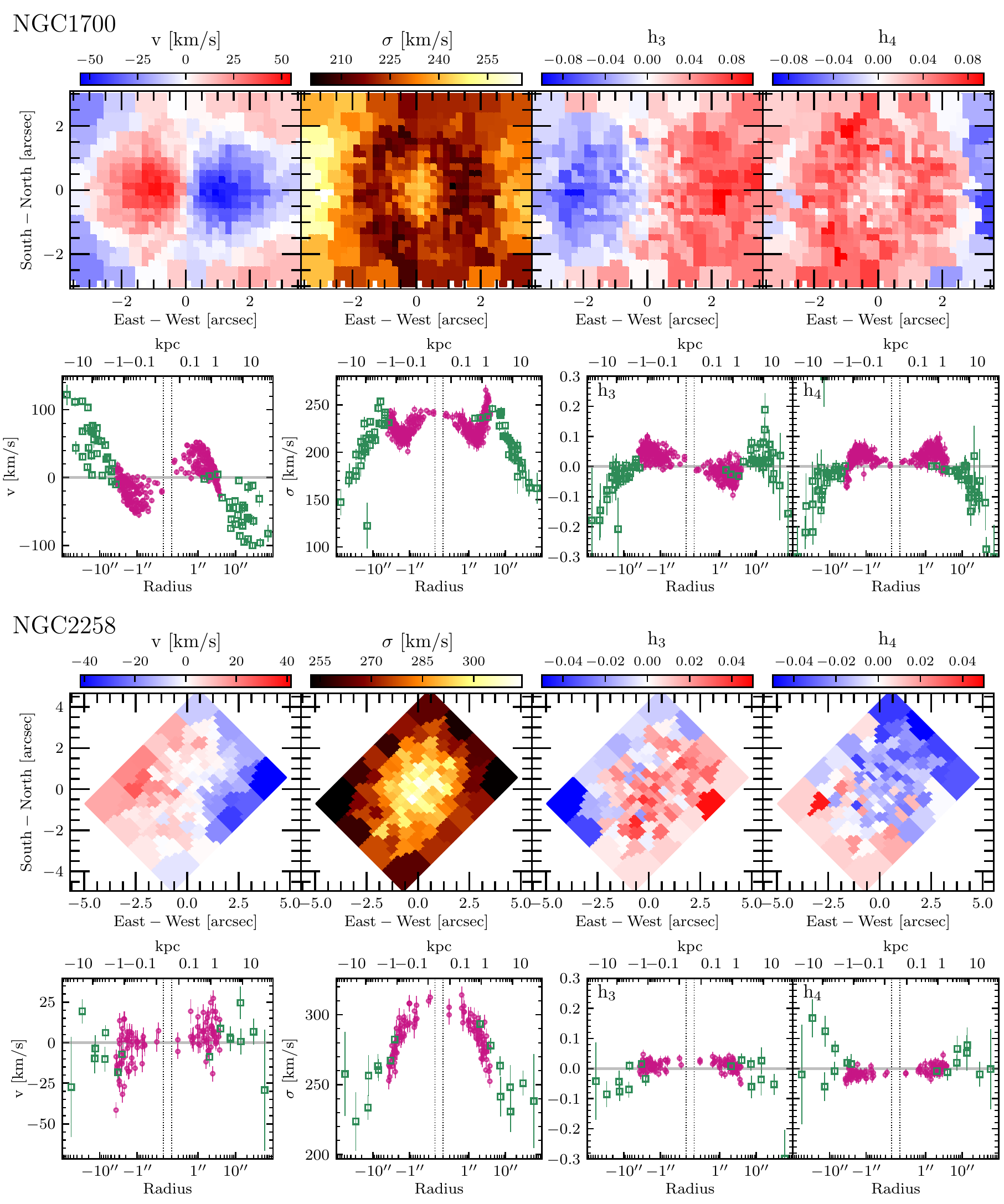} 
\caption{Same as in Fig.~\ref{fig:A1} but for NGC~1700 and NGC~2258.}
\label{fig:A8}
\end{figure*}

\begin{figure*}
    \includegraphics[width=\textwidth]{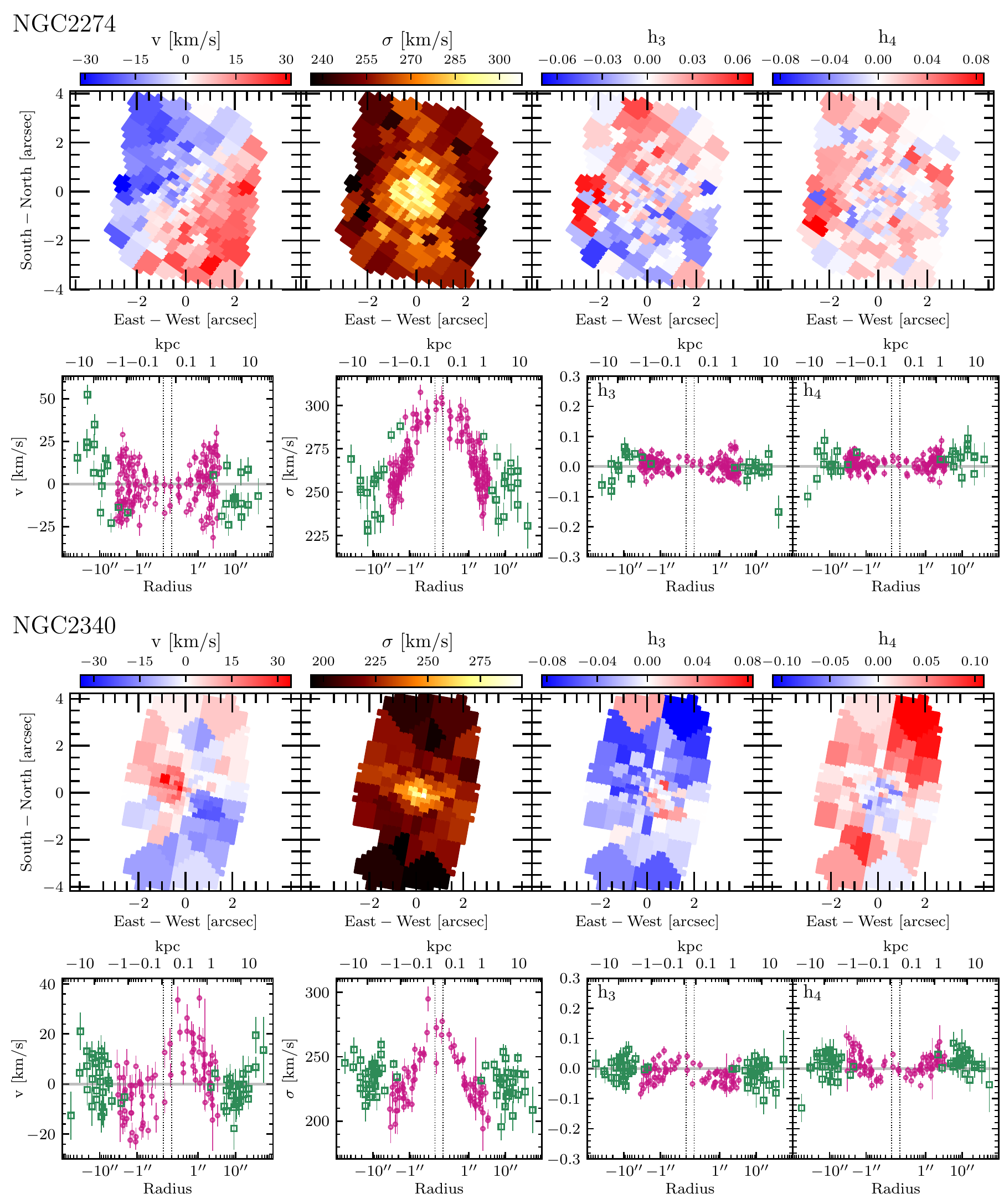} 
\caption{Same as in Fig.~\ref{fig:A1} but for NGC~2274 and NGC~2340.}
\end{figure*}

\begin{figure*}
    \includegraphics[width=\textwidth]{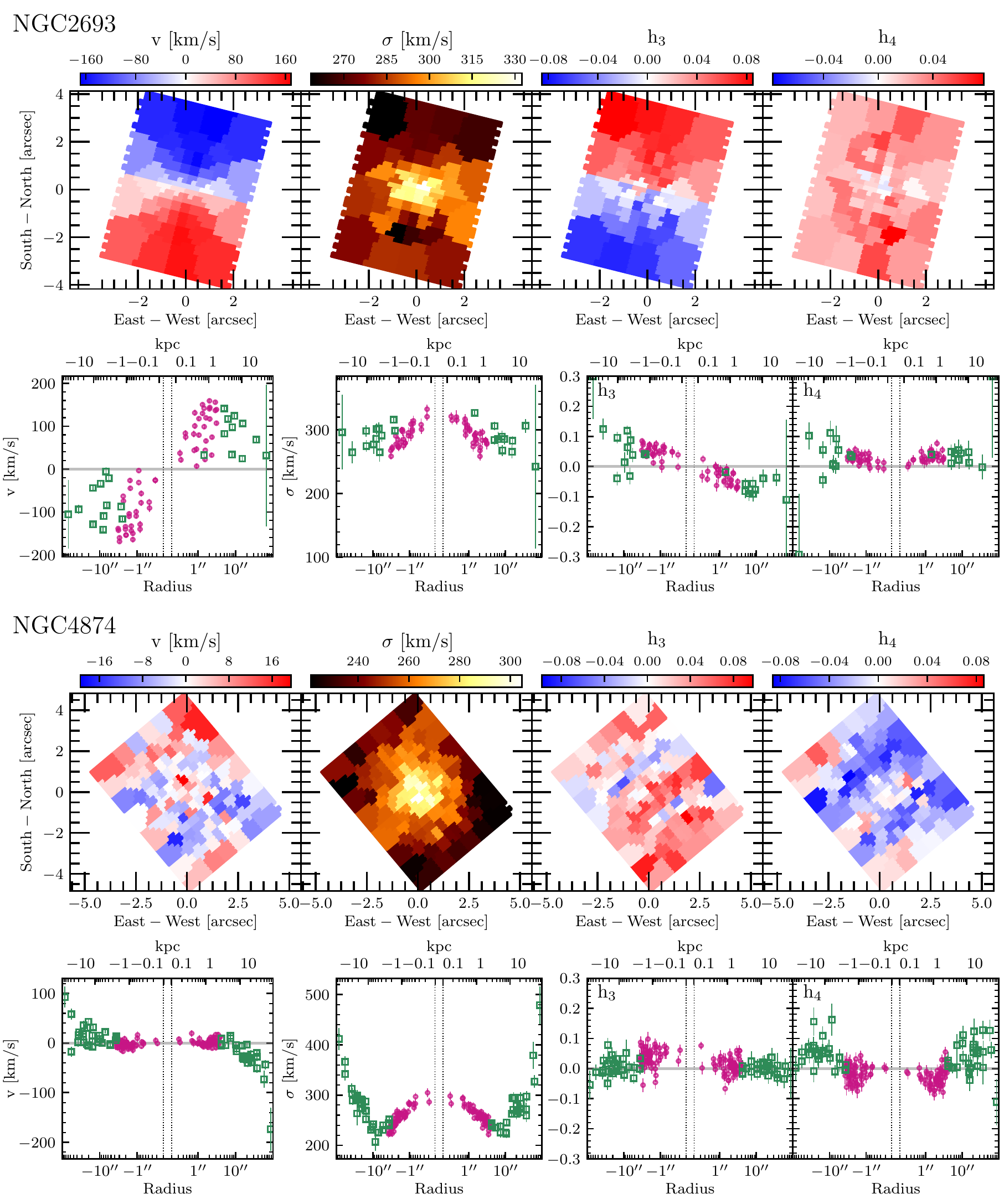} 
\caption{Same as in Fig.~\ref{fig:A1} but for NGC~2693 and NGC~4874.}
\label{fig:A10}
\end{figure*}




\bibliography{gmos}



\end{document}

%% file: galaxy_properties.tex
\startdata
NGC0057 & 76.3 & -25.75 & 40.2& 17.1 (6.31)    & 0.028 & 257 $\pm$ 2 & 0.025 & 301 $\pm$ 2 & -0.120 $\pm$ 0.006 &  41 & 2016B &12 $\times$ 1150s\\
NGC0315 & 70.3 & -26.30 & 44.3&  27.0 (9.20)   & 0.063 & 341 $\pm$ 1 & 0.027 & 325 $\pm$ 1 & -0.025 $\pm$ 0.007 &  0 &2012B & 10 $\times$ 1200s\\
NGC0410 & 71.3 & -25.90 & 35.8&  21.9 (7.57)   & 0.048 & 258 $\pm$ 1 & 0.052 & 288 $\pm$ 2 & -0.070 $\pm$ 0.006 & 35 &2016B &  7 $\times$ 1150s\\
NGC0545 & 74.0 & -25.83 & 57.2&  27.1 (9.71)   & 0.081 & 236 $\pm$ 2 & 0.034 & 236 $\pm$ 1 & -0.101 $\pm$ 0.005 & 60 &2013B &  8 $\times$ 1200s\\
NGC0547 & 71.3 & -25.90 & 98.8&  30.5 (10.55)  & 0.081 & 230 $\pm$ 2 & 0.024 & 246 $\pm$ 3 & -0.139 $\pm$ 0.008 & 94 &2016B &  8 $\times$ 1200s\\
NGC0741 & 73.9 & -26.06 & 88.0&  27.2 (9.74)   & 0.050 & 289 $\pm$ 3 & 0.037 & 274 $\pm$ 1 & -0.025 $\pm$ 0.010 & 177 &2012B &  6 $\times$ 1200s\\
NGC0777 & 72.2 & -25.94 & 148.6&  16.8 (5.89)  & 0.060 & 293 $\pm$ 2 & 0.027 & 320 $\pm$ 2 & -0.068 $\pm$ 0.004 & 148 &2015B &  8 $\times$ 1050s\\
NGC0890 & 55.6 & -25.50 & 53.7&  24.5 (6.62)   & 0.136 & 196 $\pm$ 2 & 0.027 & 206 $\pm$ 1 &  0.013 $\pm$ 0.007 & 55 &2015B &  6 $\times$ 850s\\
NGC1016 & 95.2 & -26.33 & 42.8&  20.5 (9.47)   & 0.040 & 279 $\pm$ 1 & 0.015 & 300 $\pm$ 2 & -0.069 $\pm$ 0.007 & 40 &2013B &  6 $\times$ 600s\\
NGC1060 & 67.4 & -26.00 & 74.8&  19.5 (6.38)   & 0.048 & 282 $\pm$ 2 & 0.034 & 301 $\pm$ 2 & -0.050 $\pm$ 0.008 & 74 &2016B &  7 $\times$ 1150s\\
NGC1129 & 73.9 & -26.14 & 61.7&  45.0 (16.13)  & 0.124 & 267 $\pm$ 2 & 0.350 & 228 $\pm$ 3 & -0.105 $\pm$ 0.031 & 5 &2016B &  18 $\times$ 1200s\\
NGC1453 & 56.4 & -25.67 & 30.1&  21.9 (6.00)   & 0.204 & 276 $\pm$ 1 & 0.199 & 293 $\pm$ 2 & -0.082 $\pm$ 0.011 & 20 &2015B &  6 $\times$ 850s\\
NGC1573 & 65.0 & -25.55 & 31.7&  17.2 (5.43)   & 0.056 & 270 $\pm$ 2 & 0.026 & 282 $\pm$ 2 & -0.057 $\pm$ 0.011 & 35 &2013B &  6 $\times$ 600s\\
NGC1600 & 63.8 & -25.99 & 8.8&  29.6 (9.14)    & 0.035 & 299 $\pm$ 1 & 0.045 & 337 $\pm$ 1 & -0.048 $\pm$ 0.008 & 15 &2014B &  9 $\times$ 1230s\\
NGC1700 & 54.4 & -25.60 & 90.6&  16.9 (4.45)   & 0.198 & 227 $\pm$ 1 & 0.119 & 227 $\pm$ 1 & -0.062 $\pm$ 0.005 & 90 &2015B &  13 $\times$ 850s\\
NGC2258 & 59.0 & -25.66 & 150.8&  20.2 (5.76)  & 0.071 & 258 $\pm$ 3 & 0.034 & 285 $\pm$ 1 & -0.084 $\pm$ 0.007 & 135 &2016B &  14 $\times$ 1200s\\
NGC2274 & 73.8 & -25.69 & 165.0&  18.4 (6.57)  & 0.073 & 261 $\pm$ 1 & 0.042 & 270 $\pm$ 1 & -0.077 $\pm$ 0.007 & 169 &2015B &  8 $\times$ 1050s\\
NGC2340 & 89.4 & -25.90 & 80*&  32.9 (14.27)   & 0.032 & 234 $\pm$ 1 & 0.042 & 230 $\pm$ 2 & -0.092 $\pm$ 0.013 & 170 &2012B &  11 $\times$ 1200s\\
NGC2693 & 74.4 & -25.76 & 161.3&  15.6 (5.63)  & 0.294 & 296 $\pm$ 2 & 0.337 & 291 $\pm$ 2 & -0.073 $\pm$ 0.011 & 166 &2016B &  6 $\times$ 1200s\\
NGC4874 & 102.0 & -26.18 & 40*&  38.8 (19.20)  & 0.072 & 260 $\pm$ 1 & 0.018 & 270 $\pm$ 2 & -0.110 $\pm$ 0.008 & 145 &2015A &  18 $\times$ 1220s\\
\enddata